\documentclass[aps,prb,twocolumn,superscriptaddress]{revtex4}

\usepackage{graphicx}  
\usepackage{subfigure}
\usepackage{multirow}

\usepackage{fancyhdr}
\usepackage{longtable}
\usepackage{lipsum}
\usepackage{parskip}
\usepackage[T1]{fontenc}
\usepackage{dcolumn}   
\usepackage{braket}
\usepackage{bm}        
\usepackage{amsfonts}  
\usepackage{amsmath}   
\usepackage{amssymb}   
\usepackage{natbib}
\usepackage{appendix}
\usepackage{hyperref}
\begin{document}

\title{Quantum tasks assisted by quantum noise}

\author{Chuqiao Lin}
\affiliation{Rudolf Peierls Centre for Theoretical Physics, University of Oxford, Oxford OX1 3PU, United Kingdom}

\author{Vir B. Bulchandani}
\affiliation{Princeton Center for Theoretical Science,  Princeton University, Princeton, New Jersey 08544, USA}
\affiliation{Department of Physics, Princeton University, Princeton, New Jersey 08544, USA}

\author{S. L. Sondhi}
\affiliation{Rudolf Peierls Centre for Theoretical Physics, University of Oxford, Oxford OX1 3PU, United Kingdom}

\date{June 12, 2023}

\begin{abstract}
We introduce a notion of expected utility for quantum tasks and discuss some general conditions under which this is increased by the presence of quantum noise in the underlying resource states. We apply the resulting formalism to the specific problem of playing the parity game with ground states of the random transverse-field Ising model. This demonstrates a separation in the ground-state phase diagram between regions where rational players will be ``risk-seeking'' or ``risk-averse'', depending on whether they win the game more or less often in the presence of disorder. The boundary between these regions depends non-universally on the correlation length of the disorder. Strikingly, we find that adding zero-mean, uncorrelated disorder to the transverse fields can generate a weak quantum advantage that would not exist in the absence of noise.
\end{abstract}

\maketitle 

\section{Introduction}

Entangled many-body quantum states are common in the natural world but are generically not useful for universal quantum computation. This raises the question of whether physically realistic quantum states might be used to accomplish more modest quantum tasks, that are still classically impossible and probe nontrivial features of quantum mechanics, but are easier to analyze theoretically than a universal quantum computer. This question has recently been investigated for quantum nonlocal games~\cite{Daniel_2021,daniel2022quantum,BBS2023,BBS2023TC}, i.e. cooperative games for which a set of players who share an entangled quantum state before playing will win with strictly higher probability than the best possible classical players. Many of the nonlocal games studied in the latter papers further have the unusual~\cite{Brassard2005} property of being ``scalable'': versions of these games exist for any number of players $N \geq 3$, who can attain quantum advantage by sharing $\mathcal{O}(N)$ qubit states before playing the game.

Such games are of theoretical interest because they are simple enough to be analyzed in some detail while also probing fundamental properties of entangled many-body quantum states such as contextuality\cite{Abramsky2017,cabello2021converting}, which is believed to enable the measurement-based model of universal quantum computation\cite{Rauss,howard2014contextuality}. Thus a precise quantification of which states are ``good'' resources for winning quantum nonlocal games can be seen as a first step towards the much more ambitious goal of characterizing which states are useful for universal quantum computing. It is worth emphasizing that the most popular measure of quantum entanglement, namely the entanglement entropy, is wholly inadequate for this purpose\cite{EntUseless,chen2022quantum}.

The specific question that we consider in this paper is how the ``expected utility'' of a quantum task is modified by randomness of the underlying quantum states. The expected utility for a quantum task will be defined carefully below in Section \ref{sec:QuantumTasks}; for now it can be thought of as a real number that quantifies the rate of success at the quantum task in question, with larger values corresponding to a higher rate of success. The mathematics needed to analyze this problem is standard within economics~\cite{von2007theory,pratt1964risk} but less commonly applied within physics. For simplicity, our presentation will focus on random ensembles of pure states (there are no serious difficulties in extending the discussion to mixed states). In the context of nonlocal games, our analysis can be viewed as an extension of earlier results~\cite{Daniel_2021,daniel2022quantum,BBS2023,BBS2023TC} to allow for quantum noise. We note that while previous studies have looked at the effects of specific forms of non-unitary\cite{Brassard2005} and unitary\cite{fialik2010unitary} noise on nonlocal games, the general features of this problem that we identify below do not seem to have been discussed in the literature. We will illustrate the resulting theory using the example of the parity game. 

The original perfect quantum strategy for the parity game is due to Brassard-Broadbent-Tapp (BBT)~\cite{brassard2005recasting}, building on earlier results by Mermin~\cite{mermin1990extreme} on the Greenberger-Horne-Zeilinger (GHZ) state. Their proposal involves applying a specific sequence of gates and measurements (that we call the ``BBT protocol'') to the GHZ state. A recent study involving two of the present authors systematically examined how well a set of quantum players could perform at the parity game by applying the BBT protocol to a general pure state $|\psi\rangle$, instead of the GHZ state~\cite{BBS2023}. That work focused on the case that $|\psi\rangle$ was the ground state of the transverse-field Ising model, and found that the resulting ``quantum advantage'' (measured by the difference $\Delta p$ between the quantum probability of the players winning the game and the best possible classical probability of winning) could range from an order one positive number, to a positive number exponentially small in the number of players, to a negative number, with the specific location of these regimes within the Ising phase diagram depending on the version of the parity game being considered. In what follows, we will refer to these cases as ``strong quantum advantage'', ``weak quantum advantage'', and ``no quantum advantage'' for the parity game respectively. We distinguish strong and weak quantum advantage in this way for the reason that resolving an exponentially small value of $\Delta p > 0$ will in general require a number of experimental trials that is exponentially large, which departs from the colloquial notion of quantum advantage based on polynomial-time quantum algorithms. (Our distinction between strong and weak quantum advantage is analogous to the distinction between the classical probabilistic computational complexity classes BPP and PP~\cite{Gill}.)

These results illustrate how, upon playing the parity game with states that are somewhat more physically natural (from the viewpoint of condensed matter physics) than the GHZ state, multiple qualitatively different regimes of quantum advantage become possible. An immediate question is how robust these different kinds of quantum advantage are to the presence of quantum noise, beyond the quantum fluctuations inherent in the state $|\psi\rangle$. The particular type of quantum noise of interest depends on how the state $|\psi\rangle$ is realized physically. For example, if $|\psi\rangle$ arises experimentally as the ground state of a many-body Hamiltonian $\hat{H}$, a potentially significant source of quantum noise is disorder in the couplings of $\hat{H}$. From this viewpoint, a minimal extension of our previous analysis for the parity game to allow for quantum noise consists of studying the quantum winning probability when the parity game is played with ground states of the random transverse-field Ising model (RTFIM). This will provide the central example for our study below. 

The paper is structured as follows. We first introduce a notion of expected utility for quantum tasks and discuss its response to quantum noise in the underlying resource states. We observe that the sign of this response is perturbatively determined by the Hessian of the utility function and note an analogy with the theory of risk aversion in economics~\cite{arrow1964role,pratt1964risk}. We then propose a notion of expected utility for the parity game, and illustrate how this behaves when the parity game is played with ground states of the RTFIM. We find in general that the effect of disorder on the players' probability of winning the game depends non-universally on the correlation length of the disorder, with possible singularities at the quantum critical point of the underlying TFIM. We further exhibit examples for which adding zero-mean disorder to the transverse fields generates a weak quantum advantage, despite there being no quantum advantage for the clean system. We conclude with some natural open questions.

\section{Expected utility for quantum tasks} 
\label{sec:QuantumTasks}
\subsection{General theory}
\label{sec:gentheor}
Our operational definition of a quantum task $Q$ will be any sequence of unitary gates and projective measurements applied to some (pure) quantum state $|\psi\rangle \in \mathcal{H}$, where $\mathcal{H}$ denotes the set of possible resource states. We will further assume that the effectiveness of the state $|\psi\rangle$ for performing the task $Q$ is quantified by a real-valued function $u : \mathcal{H} \to \mathbb{R}$, which we call the ``utility function'' for the task $Q$. We emphasize that $u$ can be any measurable function of the state $|\psi\rangle$ whatsoever. (Thus $u$ is less constrained than the analogous notion of ``resource measure'' in quantum resource theories\cite{Chitambar_2019}.)

Let us now suppose that $|\psi\rangle$ exhibits quantum noise; to be precise, suppose that $|\psi\rangle$ is drawn from some random ensemble of pure states $\mathcal{E}$. Then, letting $\mathbb{E}$ denote expectation values over the ensemble $\mathcal{E}$, we define the ``expected utility'' for the ensemble $\mathcal{E}$ to be the number
\begin{equation}
\label{eq:expectedutility}
U = \mathbb{E}[u(|\psi\rangle)].
\end{equation}
We note that according to this formulation of expected utility, $u$ will generally already encode a Born-rule average over possible outcomes of projective measurements on the state $|\psi\rangle$. The average $\mathbb{E}$ thus represents an additional average over noisy quantum states $|\psi\rangle$. The resulting definition of $U$ in Eq. \eqref{eq:expectedutility} seems to us the most simple-minded way of combining these two averages. However, just as for the notion of expected utility in economic theory\cite{von2007theory}, this should be regarded as a convenient choice rather than a canonical definition, and it is conceivable that for more elaborate quantum tasks involving post-selection or feedback based on measurement outcomes, an alternative definition of $U$ will be more useful.

As a more structured example that is germane to our considerations below, suppose that $|\psi\rangle$ depends smoothly on $M$ real parameters $g_1,g_2,\ldots,g_M$, and that the ensemble $\mathcal{E}$ is defined by randomness in these couplings $\mathbf{g} \in \mathbb{R}^M$. (For example, $|\psi\rangle$ might be the ground state of a disordered Hamiltonian with $M$ random couplings.) Then we can treat $u$ as a function $u : \mathbb{R}^M \to \mathbb{R}$ and by Jensen's inequality\cite{ADictionaryofStatistics} it follows that
\begin{equation} \label{eq:jensen}
    \begin{cases}
        u(\mathbb{E}[\mathbf{g}]) \leq \mathbb{E}[u(\mathbf{g})], \quad &\text{if $u$ is convex} \\
        u(\mathbb{E}[\mathbf{g}]) \geq \mathbb{E}[u(\mathbf{g})], \quad &\text{if $u$ is concave}
    \end{cases}
\end{equation}
as a function of $\mathbf{g}$. Thus if $u(\mathbf{g})$ is a convex function, disorder in $\mathbf{g}$ will never decrease the expected utility of the quantum task $Q$. In this way, quantum noise can \textit{enhance} the expected utility of a quantum task.

At the same time, the ``global'' condition that $u(\mathbf{g})$ is convex is rather restrictive and too strong to be satisfied by realistic examples, including the main example of interest below. Let us therefore, following Pratt\cite{pratt1964risk}, consider the case of perturbatively weak randomness, with $\mathbf{g} = \bar{\mathbf{g}} + \delta \mathbf{g}$, where the $\delta g_i$ are possibly correlated random variables, with zero mean $\mathbb{E}[\delta \mathbf{g}] = 0$, covariance $C_{ij} = \mathbb{E}[\delta g_i \delta g_j] = \mathcal{O}(\sigma^2)$ where $\sigma \ll 1$, and suppose that all higher moments of $\delta \mathbf{g}$ are $\mathcal{O}(\sigma^3)$ as $\sigma \to 0$. We additionally assume that $u(\mathbf{g})$ is thrice differentiable. Then by Taylor's theorem it is immediate that
\begin{equation}
\label{eq:TaylorUtility}
\mathbb{E}[u(\mathbf{g})] = u(\bar{\mathbf{g}}) + \frac{1}{2}\sum_{i,j=1}^M C_{ij}H_{ij}(\bar{\mathbf{g}}) + \mathcal{O}(\sigma^3)
\end{equation}
as $\sigma \to 0$, where we introduced the Hessian
\begin{equation}
H_{ij}(\mathbf{g}) = \frac{\partial^2 u}{\partial g_i \partial g_j}(\mathbf{g}).
\end{equation}
It is clear that the effect of small random perturbations $\delta \mathbf{g}$ is controlled by the spectrum of the Hessian $H(\bar{\mathbf{g}})$; if the latter is positive (resp. negative) definite, such perturbations will always increase (resp. decrease) the expected utility of $Q$. For a general covariance matrix $C$, this is the most general ``local'' condition that will guarantee a definite sign for the second variation 
\begin{equation}
\label{eq:intHessian}
\delta u^{(2)} = \frac{1}{2}\sum_{i,j=1}^M C_{ij} H_{ij}(\bar{\mathbf{g}}).
\end{equation} 
However, if the covariance matrix has additional structure, then weaker conditions will suffice.

For example, in the specific case of i.i.d. zero-mean noise, which we can write as $C_{ij} = \sigma^2 \delta_{ij}$, Eq. \eqref{eq:TaylorUtility} simplifies to
\begin{equation}
\label{eq:UncorrUtility}
\mathbb{E}[u(\mathbf{g})] = u(\bar{\mathbf{g}}) + \frac{1}{2}\sigma^2 \nabla^2 u(\bar{\mathbf{g}}) + \mathcal{O}(\sigma^3),
\end{equation}
so that a sufficiently small and nonzero amount of uncorrelated noise will always be helpful for accomplishing the quantum task $Q$ provided that $\nabla^2 u(\bar{\mathbf{g}}) > 0$. Thus we have identified three increasingly weak conditions under which disorder will always increase (resp. decrease) the expected utility of $Q$. 

Finally, let us suppose that the agents executing the quantum task $Q$ in question are free to choose the quantum noise level $\sigma \ll 1$ for their experimental system. Then, borrowing standard economic terminology\cite{pratt1964risk}, economically rational agents seeking to maximize their joint expected utility $u$ will be either ``risk-seeking'' or ``risk-averse'' depending on whether the sign of $\delta u^{(2)}$ is positive or negative. 

\subsection{Example: the parity game}

We now discuss a specific realization of the theory presented above in the context of the parity game\cite{mermin1990extreme,brassard2005recasting}. We first briefly recall some key properties of this game\cite{Brassard2005,BBS2023}, so as to be self-contained.

Each round of the parity game can be played by $N\geq 3$ players and a referee. At the beginning of the game, the referee gives the $j$th player a classical bit $a_j\in \{0,1\}$. The bit string $\vec{a} = (a_1,a_2,\ldots,a_N)$ is drawn uniformly randomly from the set of $2^{N-1}$ bit strings satisfying the promise $\sum_{j=1}^N a_j \mod 2 = 0$. The players may not communicate classically with one another during the course of the game, and to win the game, they must each return a classical bit $b_j\in \{0,1\}$ to the referee. The players collectively win the game if
\begin{equation} \label{eq:win_criteria}
\sum_{j=1}^N b_j \bmod 2 = \frac{\sum_{j=1}^N a_j}{2} \bmod 2
\end{equation}
and lose the game otherwise. It can be shown that a set of purely classical players win the game with a probability at most 
\begin{equation}
p \leq p_{\mathrm{cl}}^* = \frac{1}{2}+ \frac{1}{2^{\lceil N/2 \rceil}}
\end{equation}
and that there exist classical strategies saturating this bound\cite{brassard2005recasting}.

On the other hand, a set of quantum players who share the $N$-qubit GHZ state $|\psi\rangle = |\mathrm{GHZ}^+\rangle$ before playing the game, where
\begin{equation}
\label{eq:GHZ}
|\mathrm{GHZ}^{\pm}\rangle  = \frac{1}{\sqrt{2}}\left(|00 \ldots 0\rangle \pm |11\ldots 1\rangle\right),
\end{equation}
can win the game with probability one. In order to achieve this, the players apply the BBT protocol of gates and measurement $\mathcal{P}_\mathrm{BBT}$ to their shared state $|\psi\rangle$, defined as follows: First, each player acts on their qubit with an input-dependent phase gate,
\begin{equation}
    \hat{Z}^{a_j/2} = \left(\begin{array}{cc}
1 & 0 \\
0 & i^{a_j}
\end{array}\right).
\end{equation}
Second, each player rotates their basis to the $\hat{X}$ or Hadamard basis by applying the unitary
\begin{equation}
    \hat{U}=\frac{1}{\sqrt{2}}\left(\begin{array}{cc}
1 & 1 \\
1 & -1
\end{array}\right).
\end{equation}
Finally, each player measures their qubit in the $\hat{Z}$ basis, yielding a measurement outcome $(-1)^{b_j}$, and returns the bit $b_j$. We refer to Mermin\cite{mermin1990extreme} and BBT\cite{brassard2005recasting} for an explanation of why this quantum strategy always wins the game.

In previous work\cite{BBS2023}, we generalized the BBT protocol to arbitary states $|\psi\rangle$ by considering the quantum strategy $\mathcal{S} = (|\psi\rangle,\mathcal{P}_{\mathrm{BBT}})$ for playing the parity game, which consisted of applying the input-dependent protocol $\mathcal{P}_{\mathrm{BBT}}$ to an arbitrary pure state $|\psi\rangle$. There we found the explicit formula
\begin{equation} \label{eq:theorem1}
    p_\mathrm{qu}(\ket{\psi}) = \frac{1}{2}\left(1+\left|\braket{\mathrm{GHZ}^+|\psi}\right|^2-\left|\braket{\mathrm{GHZ}^-|\psi}\right|^2\right)
\end{equation}
for the quantum probability of winning the game, which attains its maximum for the GHZ state $|\psi\rangle = |\mathrm{GHZ}^+\rangle$. This result can be seen as a simplified but practically useful form of the full ``rigidity'' statement for the parity game~\cite{Werner_2001,colbeck2011quantum,mckague2010selftesting,miller2013optimal}.

For example, in previous work we computed Eq. \eqref{eq:theorem1} exactly for the $g>0$ ground state of the transverse-field Ising model
\begin{equation}
\label{eq:TFIM}
\hat{H}_{\rm TFIM} = -\sum_{j=1}^N \hat{Z}_j \hat{Z}_{j+1} - g \sum_{j=1}^N \hat{X}_j
\end{equation}
on a ring with $j \equiv j+N$, finding the expressions\cite{BBS2023}\footnote{The last line corrects a stray factor of two in a previously published formula\cite{BBS2023}.}
\begin{equation} \label{eq:explicit_pqu}
    \begin{aligned}
    p_{\mathrm{qu}}(g) & =\frac{1}{2}+\frac{1}{2}\left|\left\langle\mathrm{GHZ}^{+} \mid \psi(g)\right\rangle\right|^2 \\
    & =\frac{1}{2}+\frac{1}{2} \prod_{k>0} \cos ^2\left(\frac{\theta_k(g)-\theta_k(0^+)}{2}\right) \\
    & =\frac{1}{2}+\frac{1}{2^{\lfloor N / 2\rfloor+1}} \prod_{k>0}\left(1+\frac{1-g \cos k}{\sqrt{1+g^2-2 g \cos k}}\right)
    \end{aligned}
\end{equation}
with the allowed wavenumbers $k$ defined by
\begin{align}
    k= \begin{cases} \pm \frac{\pi}{N}, \pm \frac{3 \pi}{N}, \ldots \pm \frac{(N-1) \pi}{N}, & N \text { even, } \\ \pm \frac{\pi}{N}, \pm \frac{3 \pi}{N}, \ldots \pm \frac{(N-2) \pi}{N}, & N \text { odd,}\end{cases}
\end{align}
where $\tan{\theta_k(g)} = \sin{k}/(g-\cos{k})$ and it will be useful to define $\epsilon_k(g) = \sqrt{1+g^2-2g\cos k}$ (see discussion around Eq. \eqref{eq:bogoangle} for further details). We found that for all $0<g<1.506\ldots$, this quantum strategy exhibited a weak quantum advantage relative to the optimal classical probability of winning $p_{\mathrm{cl}}^*$, i.e. $\Delta p = p_{\mathrm{qu}}(g)-p_{\mathrm{cl}}^*$ was positive but exponentially small in $N$, with both $p_{\mathrm{qu}}(g)$ and $p_{\mathrm{cl}}^*$ themselves exponentially small corrections to the winning probability for random guessing, $p_\mathrm{r} = 1/2$. We previously referred to this competition as a ``battle of exponentials'' between the exponentially small differences $p_{\mathrm{qu}} (g)- p_\mathrm{r}$ and $p_{\mathrm{cl}}^*-p_\mathrm{r}$ .

The question now arises of how best to define the collective utility of a set of players of a nonlocal game. We emphasize from the beginning that this choice is arbitrary; as a simple example, if the players collectively receive a payoff $1$ whenever they win the game and a payoff $0$ whenever they lose the game, their expected payoff (the standard notion of utility for games\cite{von2007theory}) recovers the quantum probability of winning. Thus for the parity game and the strategy $\mathcal{S} = (|\psi\rangle,\mathcal{P}_{\rm BBT})$, this definition yields the utility function
\begin{equation}
u(|\psi\rangle) = p_{\mathrm{qu}}(|\psi\rangle).
\end{equation}

An undesirable property of this utility function is that it does not usefully capture the quantum advantage of the players relative to a set of purely classical players. A naive modification of this quantity is $u(|\psi\rangle) = \Delta p = p_{\mathrm{qu}} - p_{\mathrm{cl}}^*$, which has the convenient property that it is only positive if the strategy $\mathcal{S}$ yields a quantum advantage for the parity game. However, the exponential smallness of this quantity makes it unsuitable for directly probing weak quantum advantage as arises quite generically for physically realistic even-parity states\cite{BBS2023}. From this viewpoint, a ``useful'' measure of weak quantum advantage is one that quantifies the rate of decay of this exponential with the number of players, compared to the classical probability of winning, leading us to the definition
\begin{equation}
\label{eq:utilityfn}
u(|\psi\rangle) = \log{\left(\frac{p_{\mathrm{qu}}(|\psi\rangle)-p_\mathrm{r}}{p_{\mathrm{cl}}^* - p_\mathrm{r}}\right)},
\end{equation}
which we shall adopt throughout the remainder of this work. We emphasize that this formula can be applied to any nonlocal game and quantum strategy with $p_{\mathrm{qu}} > p_\mathrm{r}$ (there is no point in considering strategies that perform worse than random guessing). We note that when the quantum strategy $\mathcal{S}$ is optimal, the quantities $2\Delta p$ and $(p_{\mathrm{qu}}-p_r)/(p^*_{\mathrm{cl}}-p_r)$ are known as the ``bias difference" and the ``bias ratio" respectively~\cite{biasratio}; in this language, our preferred notion of utility for suboptimal quantum strategies $\mathcal{S}$ is the logarithm of a suboptimal bias ratio.

For the parity game and the quantum strategy $\mathcal{S} = (|\psi\rangle, \mathcal{P}_{\mathrm{BBT}})$, where $|\psi\rangle$ has even parity $\prod_{j=1}^N \hat{X}_j |\psi\rangle = |\psi\rangle$, this formula reduces to
\begin{equation}
\label{eq:utilityparity}
u(|\psi\rangle) = \log{\left(2^{\lceil N/2 \rceil-1} |\langle \mathrm{GHZ}^+|\psi\rangle|^2\right)},
\end{equation}
so that
\begin{equation}
-\infty \leq u(|\psi\rangle) \leq (\lceil N/2 \rceil-1) \log{2}.
\end{equation}
According to our terminology of strong and weak quantum advantage, the quantum strategy $\mathcal{S}$ can only exhibit either of these properties if the limit
\begin{equation} \label{eq:defb}
b = \lim_{N\to\infty} \frac{u(|\psi\rangle)}{N}
\end{equation}
exists and is strictly positive. The special case
\begin{equation}
b = \frac{\log 2}{2}
\end{equation}
corresponds to strong quantum advantage for the parity game, while all other cases
\begin{equation}
0 < b < \frac{\log 2}{2}
\end{equation}
correspond to weak quantum advantage.

For the specific case that $|\psi(g)\rangle$ is a ground state of the transverse-field Ising model with coupling strength $g>0$, Eq. \eqref{eq:explicit_pqu} implies that the utility Eq. \eqref{eq:utilityfn} of the transverse-field Ising ground state for playing the parity game is given by
\begin{equation}
\label{eq:defb2}
u(|\psi(g)\rangle) \sim N b(g), \quad N \to \infty,
\end{equation}
where
\begin{equation}
\label{eq:defbIsing}
b(g) = \int_0^{\pi} \frac{dk}{2\pi} \log{\left(1+ \frac{1-g\cos{k}}{\sqrt{1+g^2-2g\cos{k}}}\right)},
\end{equation}
which is monotonically decreasing with a unique zero $b(g_*) = 0$ at $g_* = 1.506...$. This function is plotted in Fig 1.

Thus the transverse-field Ising ground state can provide strong quantum advantage (as $g \to 0^+$), weak quantum advantage (for $0<g<g_*$) or no quantum advantage (for $g \geq g_*$) for the parity game, depending on the value of the coupling strength $g>0$.

\begin{figure}[t]
    \centering
    \includegraphics[width=0.5\textwidth]{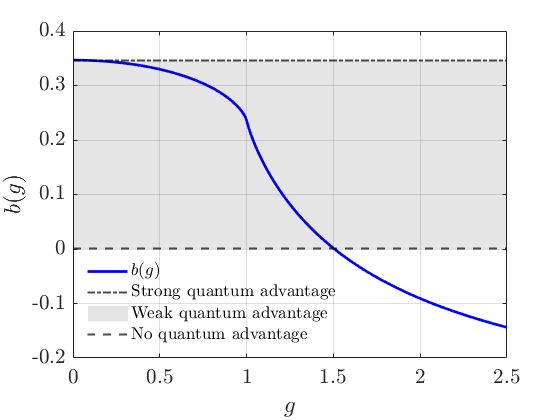}
    \caption{The quantity $b(g)$ as defined in Eq. \eqref{eq:defbIsing}. This plot shows how the ground state of the transverse-field Ising model provides a quantum advantage for the parity game that interpolates continuously between a strong quantum advantage as $g \to 0^+$ to no quantum advantage for $g \geq 1.506\ldots$, via an extended regime of weak quantum advantage. The change in slope at the critical point $g=g_c=1$ indicates a lack of smoothness at this point, which is responsible for the singular behaviour depicted in Fig. \ref{fig:RRAMaxCorr}.
    }
    \label{fig:TFIMsimu}
\end{figure}

\section{Playing the parity game with random Ising ground states}  \label{sec:results}

We now illustrate the theory developed above with the example of playing the parity game with ground states of the random transverse-field quantum Ising model,
\begin{equation}
\label{eq:rfim}
\hat{H} = -\sum_{j=1}^N \hat{Z}_j\hat{Z}_{j+1} -\sum_{j=1}^N g_j\hat{X}_j
\end{equation}
on a ring with $j \equiv j+N$, where $g_j$ denotes the strength of the transverse field at site $j$. The ground state of this system will be written as $\ket{\psi(\mathbf{g})}$, 
can be solved exactly in terms of Jordan-Wigner fermions~\cite{mbeng2020quantum}, and will always be even parity if $g_j>0$, which we henceforth assume. The vector of couplings $\mathbf{g}:=(g_1,g_2\dots,g_N)$ is a random variable whose distribution will be specified on a case-by-case basis below (thus $M=N$ in the notation of Section \ref{sec:gentheor}). Our goal will be to understand the utility Eq. \eqref{eq:utilityparity} of the state $|\psi(\mathbf{g})\rangle$ for playing the parity game, which we denote
\begin{equation}
u(\mathbf{g}) = \log{\left(2^{\lceil N/2 \rceil-1} |\langle \mathrm{GHZ}^+|\psi(\mathbf{g})\rangle|^2\right)}
\end{equation}

It will be useful to reserve the notation $\chi(g)$ for the specific case $g_1 =g_2 =\ldots=g_N =g$, i.e.
\begin{equation} 
    \chi(g):= u(g,g,\ldots,g).
\end{equation}
In terms of the Bogoliubov angles $\theta_k(g)$, we can write this function explicitly as
\begin{align}
\nonumber
\chi(g) = &(\lceil N/2 \rceil - 1) \log{2} \\
\label{eq:chi}
+ &2 \sum_{k>0} \log{\cos\left(\frac{\theta_k(g) - \theta_k(0^+)}{2}\right)}.
\end{align}
\subsection{Perfectly correlated disorder}
\label{sec:unipert}

The simplest limit to consider is perfectly correlated disorder, with $g_1=g_2 =\ldots = g_N = g$ and $g>0$ drawn from some probability distribution with small variance $\sigma^2 \ll 1$. In this case, we have
\begin{equation}
\mathbb{E}[u(\mathbf{g})] \approx \chi(\bar{g}) + \frac{\sigma^2}{2} \chi''(\bar{g}).
\end{equation}
Thus to determine the perturbative effect of disorder it suffices to compute 
\begin{equation}
\chi''(g) = \sum_{i,j=1}^N \frac{\partial^2 u}{\partial g_i \partial g_j}(g,g,\ldots,g).
\end{equation}
From Eq. \eqref{eq:chi}, we find that
\begin{equation} \label{eq:d1chi}
    \begin{aligned}
        \chi'(g) &= -\sum_{k>0} f_k(g)
    \end{aligned}
\end{equation}
and
\begin{equation} \label{eq:d2chi}
    \begin{aligned}
        \chi''(g) &= \sum_{k>0} \left[\frac{2f_k(g) \cos{\theta_k(g)}}{\epsilon_k(g)}-\frac{f_k^2(g) }{2}-\frac{\sin^2{\theta_k(g)}}{2\epsilon_k^2(g)}\right],
    \end{aligned}
\end{equation}

where we defined
\begin{equation} \label{eq:fk}
    f_k(g) := -\frac{1}{\epsilon_k(g)} \tan \left(\frac{\theta_k(g)-\theta_k(0^+)}{2}\right) \sin \theta_k(g).
\end{equation}

\begin{figure}[t]
    \centering
    \includegraphics[width=0.5\textwidth]{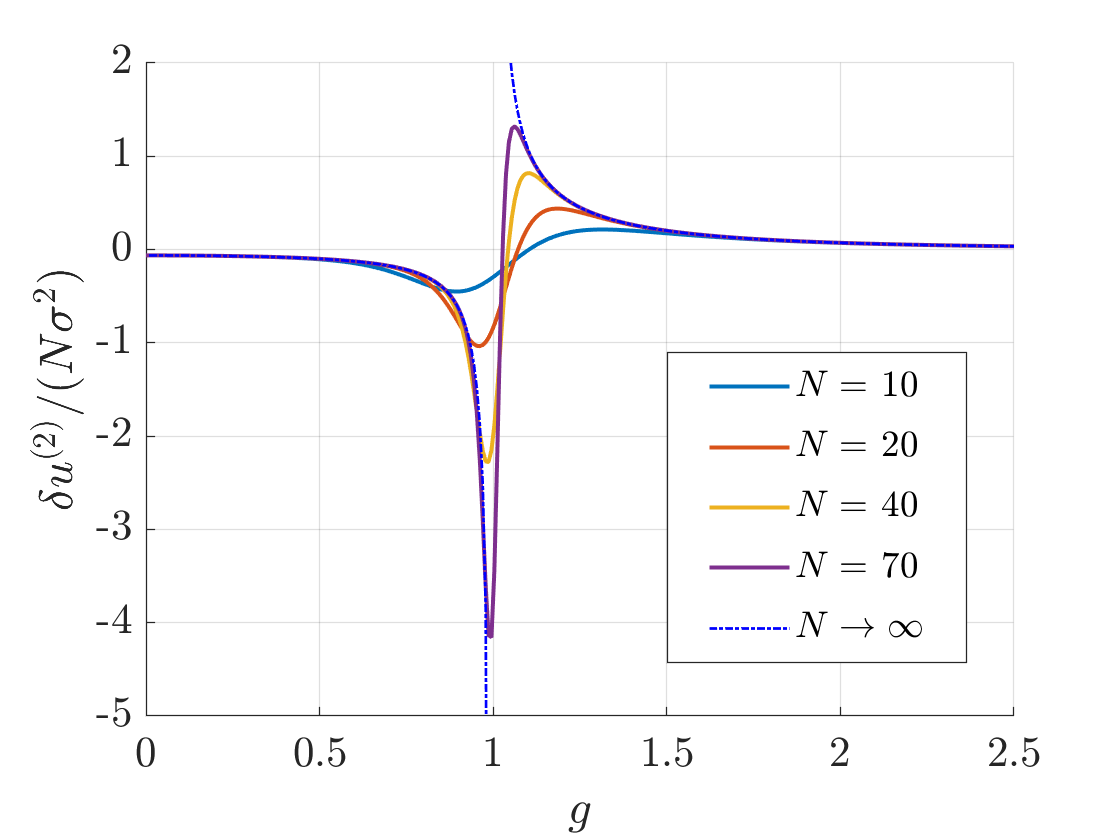}
    \caption{The second variation of utility Eq. \eqref{eq:SecondVarMaxCorr} for ground states of the random transverse-field Ising model with perfectly correlated disorder, as a function of the mean transverse-field strength $\bar{g}_i=g$ and the system size $N$. This exhibits a clear divergence at the Ising critical point $g=g_c=1$, which can be seen as a phase transition between regimes where rational players will be risk-averse (in the ferromagnetic phase) and risk-seeking (in the paramagnetic phase) respectively.}
    \label{fig:RRAMaxCorr}
\end{figure}

The rescaled second variation $\delta u^{(2)}$ in this case is given by
\begin{equation} \label{eq:SecondVarMaxCorr}
    \frac{\delta u^{(2)}}{N\sigma^2} = \frac{1}{2N}\chi''(g).
\end{equation}
Eq. \eqref{eq:SecondVarMaxCorr} is plotted in Fig. \ref{fig:RRAMaxCorr}. A striking conclusion from this plot is that rational players will \emph{prefer} weak disorder in the paramagnetic phase, because it enhances their expected probability of winning. Meanwhile, they will avoid weak disorder in the ferromagnetic phase, because this diminishes their expected probability of winning. In the large-system limit\footnote{Note that one cannot take the ordinary thermodynamic limit, because spontaneous symmetry breaking tends to eliminate quantum advantage for nonlocal games\cite{BBS2023}.} as $N \to \infty$, these two regimes are cleanly separated by a divergence in the rescaled second variation that lies precisely at the Ising critical point $g=g_c=1$. A more detailed asymptotic analysis reveals that this divergence is independent of the order of limits: the rescaled second variation diverges whether one lets $g \to g_c$ having already taken the large-system limit as $N \to \infty$, or lets $N \to \infty$ while holding $g=g_c$ fixed (see Appendix \ref{ap:scaling}).


\subsection{Uncorrelated (i.i.d.) disorder} \label{sec:normalpert}

\begin{figure}[t]
     \centering
     \begin{subfigure}
         \centering
         \includegraphics[width=0.5\textwidth]{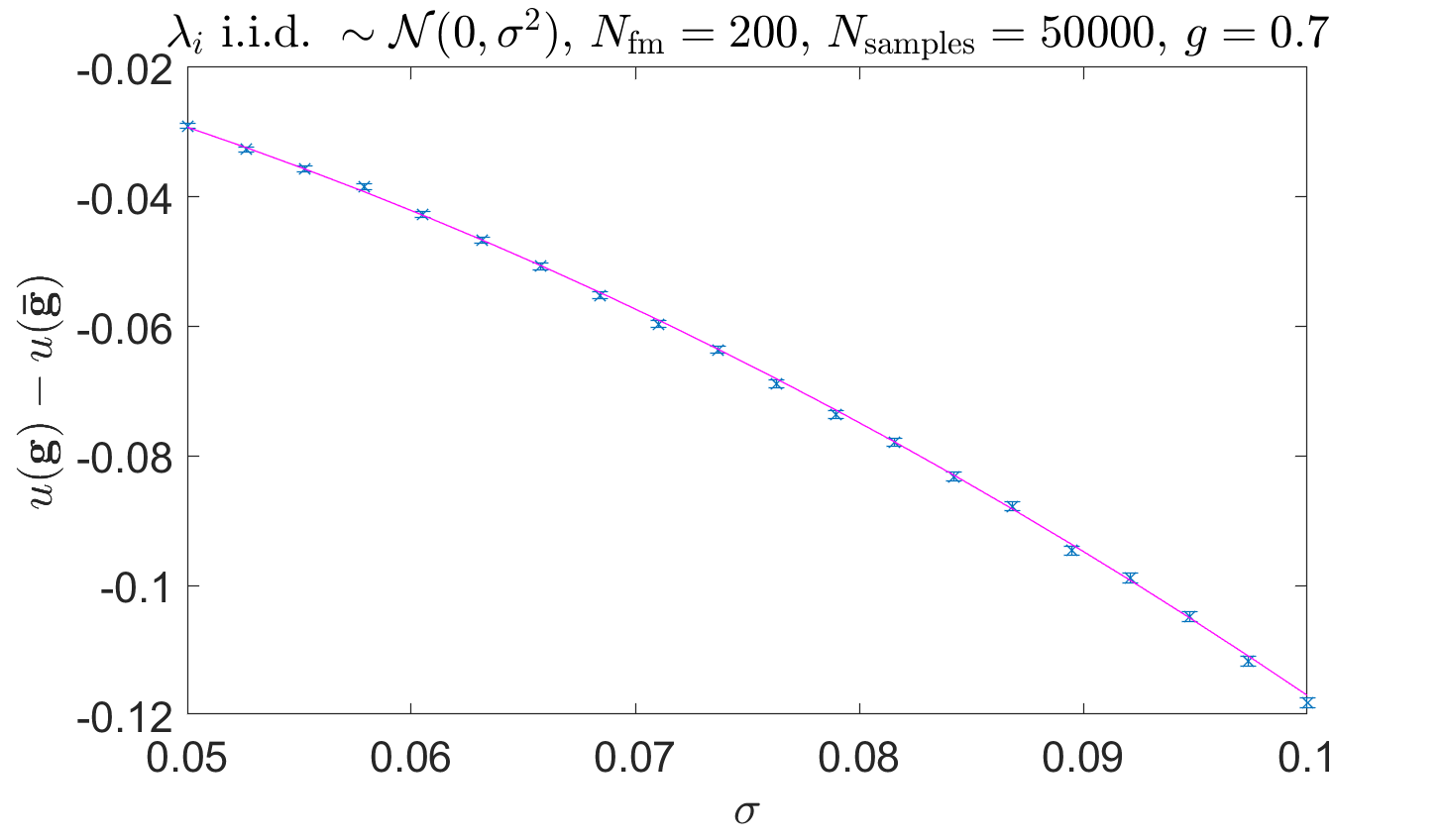}
     \end{subfigure}
     \hfill
     \begin{subfigure}
         \centering
         \includegraphics[width=0.5\textwidth]{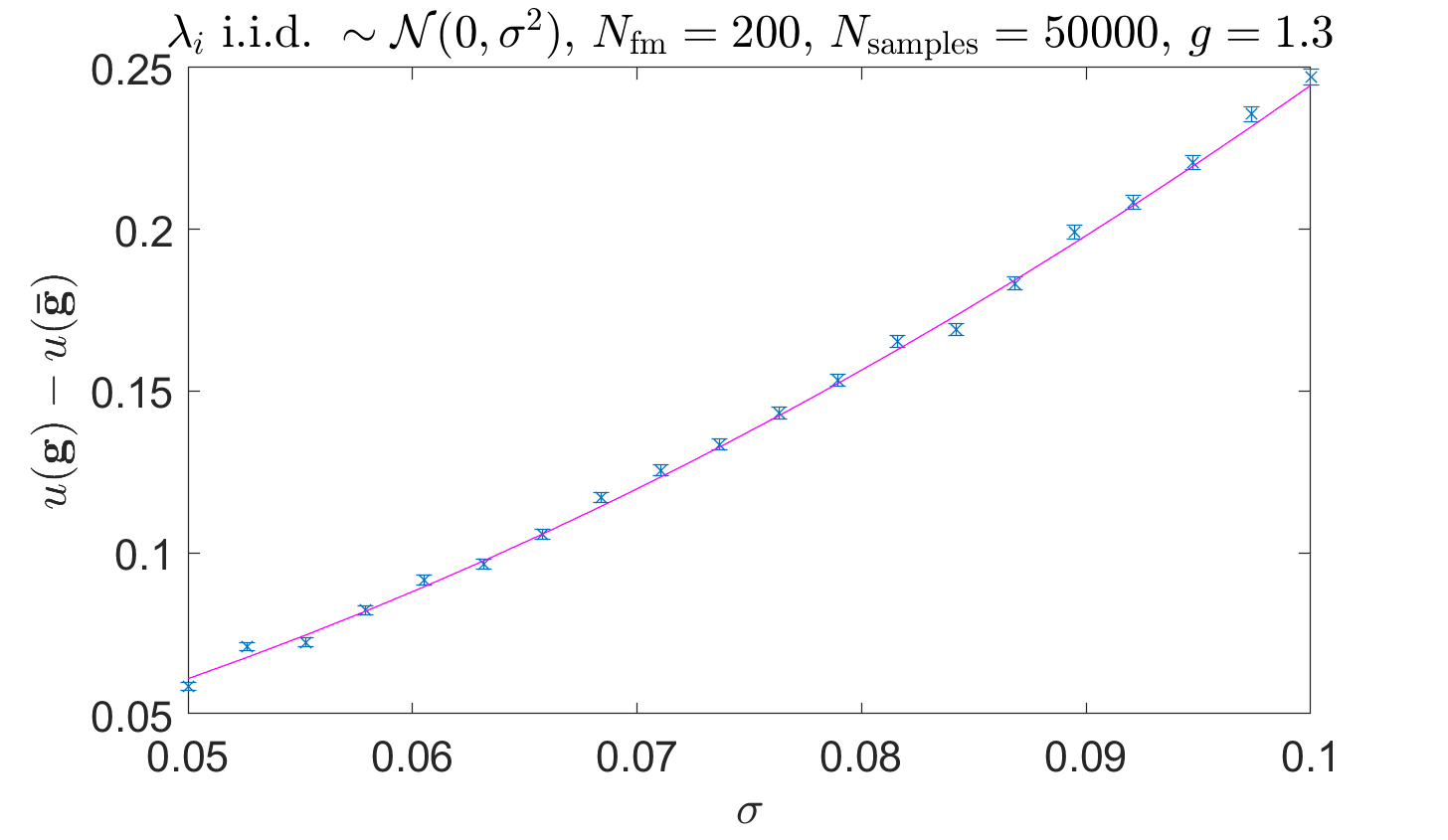}
     \end{subfigure}
       \caption{
       Leading behaviour of the expected utility Eq. \eqref{eq:uncorrRTFIM} for the RTFIM in the limit of weak and uncorrelated Gaussian disorder. The precise distributions that $\mathbf{g}$ is sampled from are defined in the figure titles. Each blue datapoint is calculated from 50000 samples at the appropriate disorder strength $\sigma$, with error bars representing the standard error of the sample mean of $u(\mathbf{g}) - u(\bar{\mathbf{g}})$. The magenta line depicts the analytical prediction $\delta u^{(2)} = \frac{\sigma^2}{2}\nabla^2u(\mathbf{g})$ for the second variation of the expected utility predicted by Eq. \eqref{eq:nabla_u}.} 
     \label{fig:normalpert}
\end{figure}

\begin{figure}[t]
    \centering
    \includegraphics[width=0.5\textwidth]{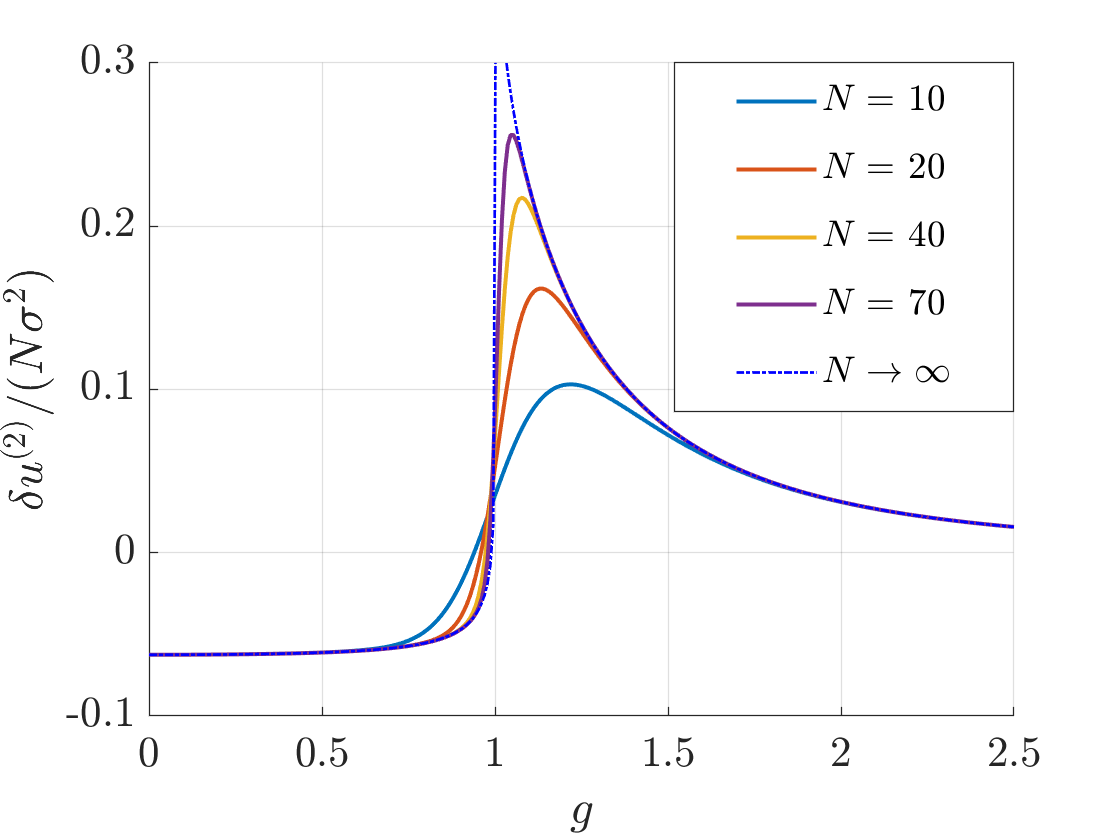}
    \caption{Eq. \eqref{eq:nabla_u} plotted as a function of the mean transverse-field strength $\bar{g}_i=g$ and the system size $N$. For finite values of $N$, there is a crossover from a regime where the players will be risk-averse, that includes nearly all of the ferromagnetic phase, to a regime where the players will be risk-seeking, which includes the entire paramagnetic phase. This crossover occurs at a transverse-field strength $g \approx 0.9902... <1$ in the large-system limit as $N \to \infty$, and is accompanied by a divergence at the Ising critical point $g=g_c=1$.}
    \label{fig:normalpertspectra}
\end{figure}

We next consider the case of i.i.d. couplings $g_i$ drawn from a distribution with $\sigma \ll \bar{g}$, so that $\delta g_i = g_i - \bar{g}$ satisfies
\begin{equation}
\label{eq:uncorrdis}
\mathbb{E}[\delta g_i] = 0, \quad \mathbb{E}[\delta g_i \delta g_j] = \sigma^2 \delta_{ij}.
\end{equation}

Then by Eq. \eqref{eq:UncorrUtility} we have
\begin{equation}
\label{eq:uncorrRTFIM}
\mathbb{E}[u(\mathbf{g})] \approx \chi(\bar{g}) + \frac{1}{2}\sigma^2 \nabla^2 u(\bar{\mathbf{g}}) 
\end{equation}

and
\begin{equation} \label{eq:nabla_u}
\begin{aligned}
   & \nabla^2u(\bar{\mathbf{g}}) = \frac{2}{N}\sum_{p_1>0, \, p_2>0} \frac{1}{(\epsilon_{p_1}+\epsilon_{p_2})^2}\left[-\epsilon_{p_1}\epsilon_{p_2}f_{p_1}f_{p_2} + \right. \\
    &  \left.  (\epsilon_{p_1}+\epsilon_{p_2})\left(f_{p_2}\cos \theta_{p_1} + f_{p_1}\cos \theta_{p_2}\right) + \left(\cos \theta_{p_1} \cos \theta_{p_2} -1\right) \right]
\end{aligned}
\end{equation}
(see Appendix \ref{ap:2ndpt} for details). This formula is verified against numerical simulations in Fig. \ref{fig:normalpert}. 

The behaviour of the rescaled second variation
\begin{equation}
\frac{\delta u^{(2)}}{N\sigma^2} = \frac{1}{2N}\nabla^2 u(\bar{\mathbf{g}})
\end{equation}
as $N \to \infty$ is depicted for a wider range of values of the mean transverse-field strength $\bar{g}$ in Fig. \ref{fig:normalpertspectra}. As for the case of perfectly correlated disorder, we find that rational players will eschew disorder in most (>$99\%$) of the ferromagnetic phase and prefer disorder in the paramagnetic phase, since the Laplacian is negative (resp. positive) in these two regimes. Similarly, the Laplacian diverges at the Ising critical point $g=g_c$. However, in contrast to the perfectly correlated case, the Laplacian changes sign at a value of $g$ that differs from the Ising critical point, $g \approx 0.9902 <g_c$ in the large-system limit. We obtain this value by approximating the sum Eq. \eqref{eq:nabla_u} by an integral, which we evaluate numerically; this integral yields the blue dashed line in Fig. \ref{fig:normalpertspectra}.

\begin{figure}[t]
     \begin{subfigure}
         \centering
         \includegraphics[width=0.45\textwidth]{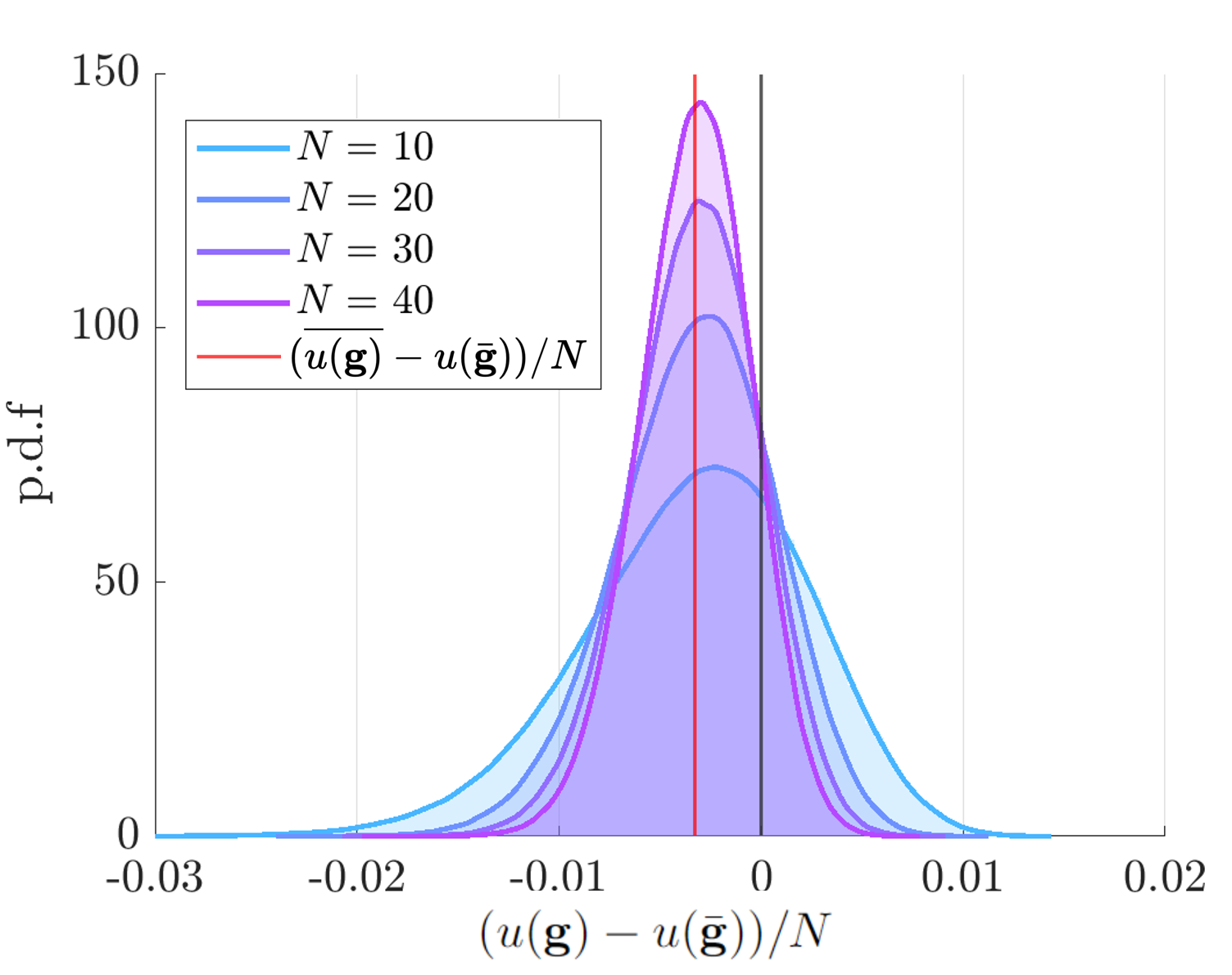}
     \end{subfigure}
     \hfill
     \begin{subfigure}
         \centering
         \includegraphics[width=0.45\textwidth]{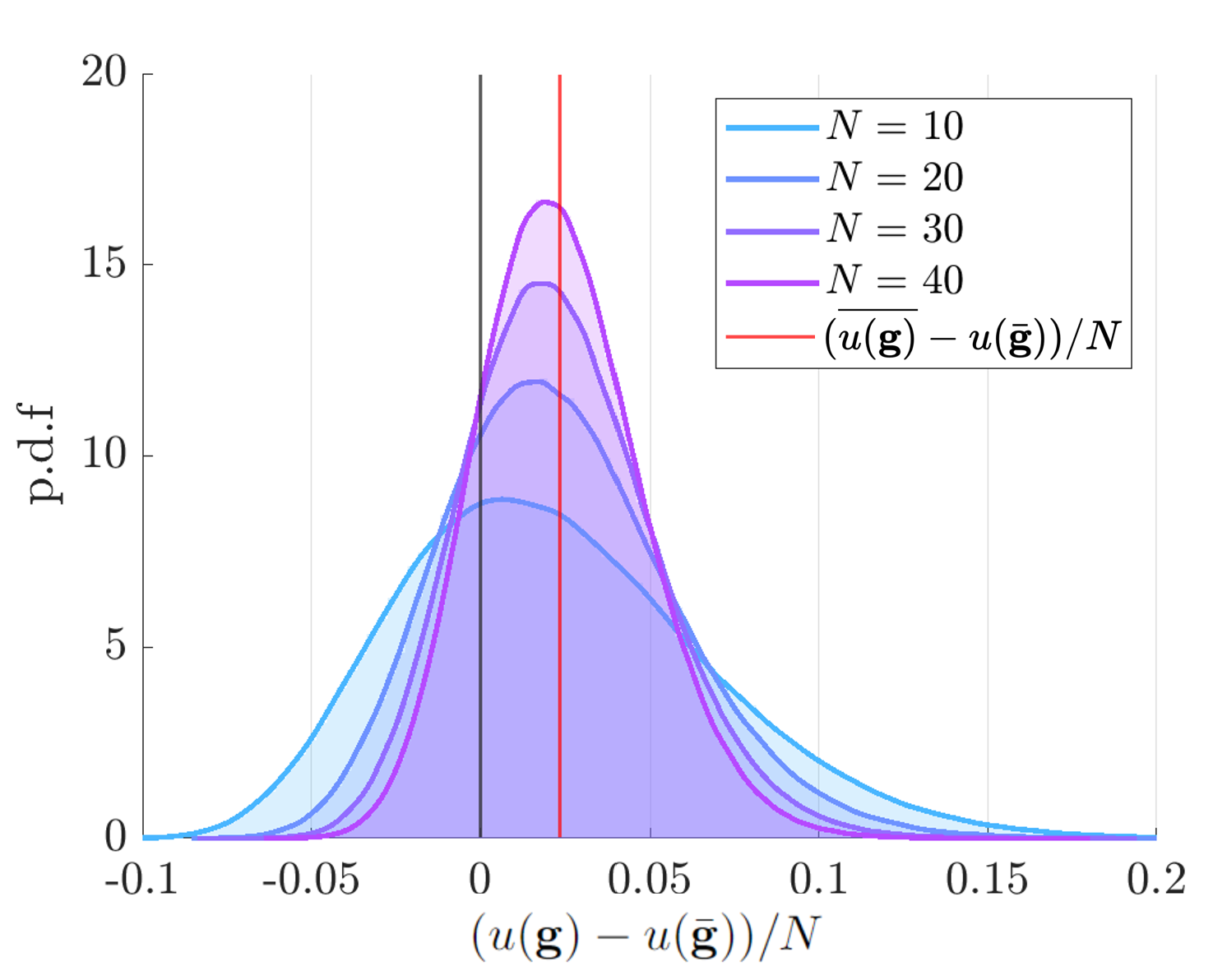}
     \end{subfigure}
     \caption{
     Histograms of the quantity $(u(\mathbf{g})-u(\bar{\mathbf{g}}))/N$ for the RTFIM with weak and uncorrelated disorder as in Eq. \eqref{eq:uncorrdis}. For each $N$, we sample $10^6$ disorder realizations $\mathbf{g}$ uniformly from the interval $[\bar{g}-W/2,\bar{g}+W/2]$, where $W$ denotes the width of the disorder (in terms of which the standard deviation  $\sigma = \frac{W}{2\sqrt{3}}$). In the top figure, $\bar g = 0.5$ and $W = 0.8$, whereas in the bottom figure, $\bar g=1.6$ and $W = 2$. These plots demonstrate an extensive shift in the expected utility $\mathbb{E}[u(\mathbf{g})]$ compared to $u(\bar{\mathbf{g}})$ as the disorder strength is increased. This shift is negative in the ferromagnetic phase (top) and positive in the paramagnetic phase (bottom). We plot the difference to systematically eliminate small-system deviations from extensivity. In the lower plot, $b(\mathbf{g}) = u(\mathbf{g})/N$ is already well-converged in $N$ for $N=40$, and we find that $b(\bar{\mathbf{g}}) = -0.022... < 0$ while $\mathbb{E}[b(\mathbf{g})] = 0.001... > 0$, i.e. disorder has induced a weak quantum advantage on average, despite the mean couplings being unchanged.
     \label{fig:RTFIMsimu}}
\end{figure}

To clarify the meaning and significance of these results, it is helpful to study the histograms of $u(\mathbf{g})-u(\bar{\mathbf{g}})$ while keeping $\sigma$ fixed and varying the number of qubits $N$. Such histograms are depicted in Fig. \ref{fig:RTFIMsimu}. Quite strikingly, we see that as $N$ is varied, there is an extensive shift in the expected utility corresponding to extensivity of $\nabla^2 u(\bar{\mathbf{g}})$, which can be deduced from Eq. \eqref{eq:nabla_u}. According to Mermin's formulation of the parity game\cite{mermin1990extreme}, this can be interpreted as an exponentially large change in the average number of satisfied GHZ stabilizers for the state $|\psi(\mathbf{g})\rangle$. Following our discussion in the previous section, this change can be negative (deep in the ferromagnetic phase) or positive (deep in the paramagnetic phase). In the paramagnetic phase, the lower panel of Fig. \ref{fig:RTFIMsimu} demonstrates an even more dramatic effect that can occur for certain parameter values, whereby the introduction of disorder tips the system from a regime of no quantum advantage to a regime of weak quantum advantage for the parity game. We have chosen to illustrate this effect away from the immediate vicinity of $g_*$ where the change in $b$ is relatively weak; by moving $\bar{g}$ closer to $g_*$, this effect can be amplified. For example, at $\bar{g}=1.55$, with the same system size and disorder distribution, we find that $b(\bar{\mathbf{g}}) = -0.010... < 0$ while $\mathbb{E}[b(\mathbf{g})] = 0.015... > 0$, so that the disorder-averaged value of $b$ is roughly an order of magnitude larger than for the example depicted in Fig. \ref{fig:RTFIMsimu}.

Finally, we note that the shape of the distributions of $u(\mathbf{g})$ becomes qualitatively normal as $N$ increases, which is consistent with the intuition that the quantum winning probability for the RTFIM ground state effectively computes the determinant of a large random matrix (see Eq. \eqref{eq:RTFIMOverlap}), and should therefore be asymptotically lognormally distributed as $N \to \infty$.

\subsection{Partially correlated disorder} 

\begin{figure}[t]
    \centering
    \includegraphics[width=0.5\textwidth]{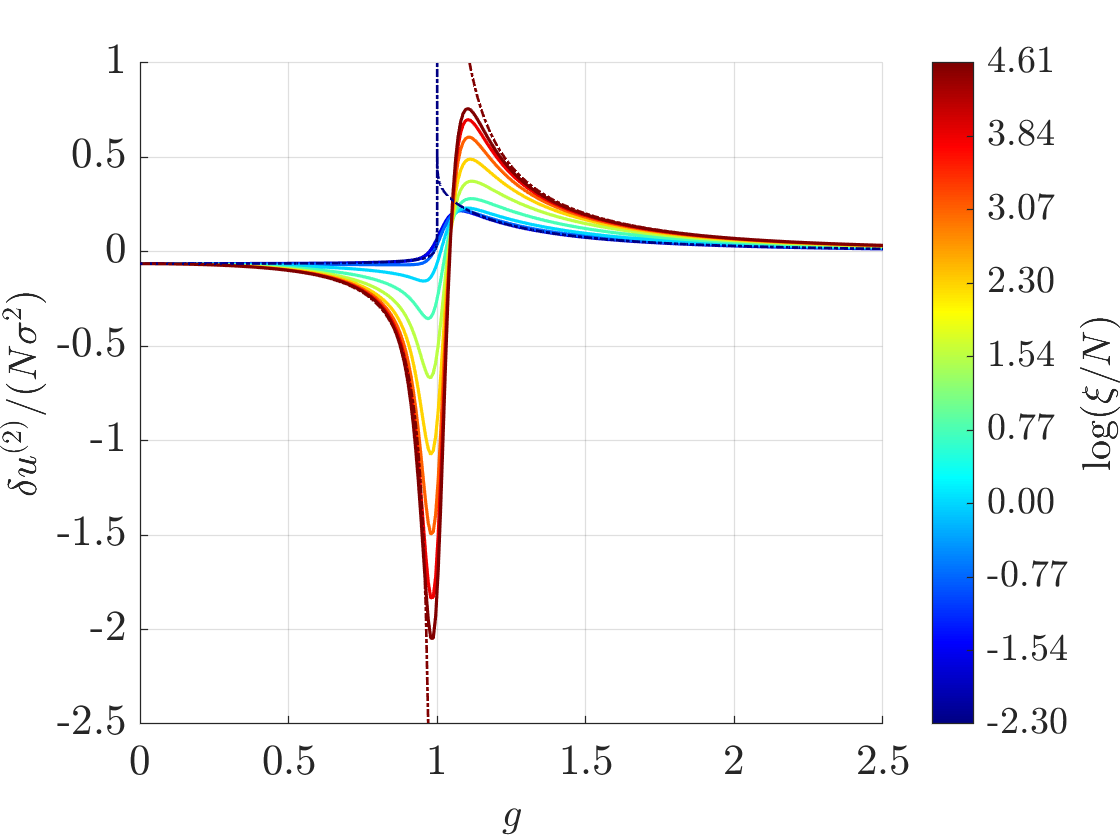}
    \caption{The second variation of the expected utility for partially correlated disorder. The colors quantify the correlation length in units of the system size $N=40$, as measured by the quantity $\log (\xi/N)$. Thus the blue curves correspond to the limit of uncorrelated disorder $\xi \ll N$, which recovers the results of Section \ref{sec:normalpert}, while the red curves correspond to the limit of highly correlated disorder $\xi \gg N$ that is discussed in Section \ref{sec:unipert}. This plot further demonstrates a smooth interpolation between the qualitative behaviours in the perfectly correlated limit (blue dotted line) and the perfectly uncorrelated limit (red dotted line) as the correlation length $\xi$ is varied. 
    }
    \label{fig:finitecorrelation}
\end{figure}

In Sections \ref{sec:unipert} and \ref{sec:normalpert} above, we considered the limits of perfectly correlated and perfectly uncorrelated disorder in the random transverse-field Ising model. We finally turn to the intermediate and physically relevant case of disorder with a finite correlation length $\xi$. Thus we suppose that the small random perturbations $\{\delta g_j\}$ satisfy
\begin{equation} \label{eq:finitecorrelationg}
    \mathbb{E}[\delta g_j] = 0,\quad \mathbb{E}[\delta g_j \delta g_l] = \sigma^2 e^{-\frac{|j-l|}{\xi}}
\end{equation}
with $\sigma \ll 1$. In order to probe the effect of the correlation length on the expected utility, we again consider the rescaled second variation of the expected utility $\frac{\delta u^{(2)}(\mathbf{g})}{N\sigma^2}$, with $\delta u^{(2)}$ obtained exactly from Eq. \eqref{eq:relationdpdu}. We find that for a fixed system size $N$, this interpolates smoothly between the behaviours identified above in the limits of perfectly correlated and perfectly uncorrelated disorder, as shown in Figure \ref{fig:finitecorrelation}. 

We can gain some insight into this behaviour through the following intuitive argument; if $\xi = \mathcal{O}(N^0)$ (the most physical case) then the system will resemble the limit of perfectly uncorrelated disorder on length scales $\gg \xi$ as $N \to \infty$, and should thus recover the qualitative behaviour observed in Sec. \ref{sec:normalpert}. Conversely, if $\xi$ grows faster than $N$, e.g. as $\xi = N^2$, then the separation of scales $N \ll \xi$ implies that the disorder will appear perfectly correlated at the system length scale $N$, recovering the qualitative behaviour observed in \ref{sec:unipert}. These intuitions are consistent with the behaviour observed in Fig. \ref{fig:finitecorrelation} and with numerical simulations for multiple values of $N$ (not shown), and can be justified more systematically from Eq. \eqref{eq:intHessian}.

\section{Conclusion}
We have developed a theoretical framework to quantify the effect of quantum noise on the successful execution of quantum tasks, and pointed out the similarities between this formalism and the theory of risk aversion in economics\cite{arrow1964role,pratt1964risk}. We illustrated this formalism with the specific example of playing the parity game with ground states of the random transverse-field Ising model, with the players free to specify the variance $\sigma^2$ of the disorder in the transverse fields. We found that in the regime of perturbatively weak disorder strength $\sigma^2 \ll 1$, the players were risk-averse over a region approximating the ferromagnetic phase, and risk-seeking over a region approximating the paramagnetic phase, with the boundary between these regions depending non-universally on the correlation length of the disorder.

A nontrivial prediction from our analysis is that the sensitivity of the quantum winning probability to disorder will diverge at the quantum critical point for disorder that is either perfectly correlated ($\xi/N \to \infty$) or perfectly uncorrelated ($\xi/N \to 0$) in the large-system limit as $N \to \infty$. This is a much more direct diagnostic of quantum criticality than the quantum winning probability itself~\cite{BBS2023}, which is continuous at the quantum critical point $g=1$ and only loses quantum advantage at $g \approx 1.5$, raising the question of how far such divergences are a universal feature of playing nonlocal games with ground states of local Hamiltonians. For example, the effects that we observe might be robust to improving\cite{miller2013optimal} the Brassard-Broadbent-Tapp protocol with single-qubit unitaries, by virtue of the divergent correlation length at criticality.

Another nontrivial effect that we have identified above is the possibility for disorder with zero mean and non-zero variance $\sigma^2 > 0$ to \emph{increase} the expected degree of weak quantum advantage for the parity game, to the point of generating a quantum advantage that would be entirely absent in the clean limit $\sigma^2=0$. It seems surprising that there are circumstances in which noise can engender a quantum advantage, given that noise is usually regarded as the bane of fault-tolerant quantum computation~\cite{Preskill_2018} and that depolarizing noise can make quantum systems easier to simulate classically~\cite{Aharonov_2023}. A few recent works have nevertheless argued that quantum noise can be beneficial within specific contexts~\cite{Du_2021,domingo2023taking}. 

In future work, it would be desirable to understand more systematically the circumstances in which quantum noise is helpful for accomplishing a given quantum task. It would similarly be interesting to find additional examples of quantum tasks that are enhanced by physically natural realizations of quantum noise, and to understand whether in such physical cases, noise can ever generate a strong quantum advantage of the type that could be discerned experimentally in a large system.

\section{Acknowledgments}

We thank F.J. Burnell for collaborations on related topics. V.B.B. is supported by a fellowship at the Princeton Center for Theoretical Science and thanks D.S. Borgnia, S.J. Garratt and A. Natarajan for helpful discussions, R.A. Bulchandani for introducing him to expected utility theory, and the Simons Institute for the Theory of Computing for their hospitality during the completion of this work. S.L.S. was supported by a Leverhulme Trust International Professorship, Grant Number LIP-202-014. For the purpose of Open Access, the author has applied a CC BY public copyright license to any Author Accepted Manuscript version arising from this submission.

\bibliography{ref}
\onecolumngrid
\appendix

\section{Quantum probability of winning the parity game with random transverse-field Ising ground states} \label{ap:diag}

In this Appendix, we show how to compute the quantum probability of winning the parity game with the quantum strategy $\mathcal{S} = (|\psi\rangle,\mathcal{P}_{\mathrm{BBT}})$, Eq. \eqref{eq:theorem1}, where $|\psi\rangle$ is the ground state of the random transverse-field Ising model Eq. \eqref{eq:rfim}.

\subsection{Conventions for TFIM diagonalization}
Since we will need them later, we first fix conventions for the standard diagonalization of the clean TFIM Eq. \eqref{eq:TFIM} via Jordan-Wigner transformation to free fermions~\cite{Damski_2013,mbeng2020quantum}. For simplicity, we will always assume periodic boundary conditions for the spin degrees of freedom.

The Jordan-Wigner transformation is
\begin{equation} \label{eq:jw}
    \hat{X}_i=1-2 \hat{c}_i^{\dagger} \hat{c}_i, \quad \hat{Z}_i=\left(\hat{c}_i+\hat{c}_i^{\dagger}\right) \prod_{j<i}\left(1-2 \hat{c}_j^{\dagger} \hat{c}_j\right)
\end{equation}
and we impose boundary conditions  $\hat{c}_{N+1} = -\hat{P}\hat{c}_1$, where $\hat{P}= \prod_{i=1}^N \hat{X}_i$ denotes the fermion parity operator. We will restrict to the even-parity sector, which includes\cite{Damski_2013} the ground state for $g>0$. A convenient choice~\cite{Damski_2013} for Fourier transformed fermions is
\begin{equation}
    \hat{c}_j =\frac{e^{-i\pi/4 }}{\sqrt{N}} \sum_k e^{i k j} \hat{c}_k
\end{equation}
with
\begin{align} 
\label{eq:wavenos}
    k= \begin{cases} \pm \frac{\pi}{N}, \pm \frac{3 \pi}{N}, \ldots \pm \frac{(N-1) \pi}{N}, & N \text { even, } \\ \pm \frac{\pi}{N}, \pm \frac{3 \pi}{N}, \ldots \pm \frac{(N-2) \pi}{N},\pi & N \text { odd, }\end{cases}
\end{align}
by anti-periodic boundary conditions. The TFIM can be diagonalized by applying the Bogoliubov transformation
\begin{equation}
    \hat{c}_k=\cos \left(\frac{\theta_k}{2}\right) \hat{\gamma}_k-\sin \left(\frac{\theta_k}{2}\right) \hat{\gamma}_{-k}^{\dagger},
\end{equation}
where the Bogoliubov angles $\theta_k$ are given by 
\begin{equation}
\label{eq:bogoangle}
\sin \theta_k = \frac{\sin k}{\epsilon_k}, \quad \cos \theta_k = \frac{g-\cos k}{\epsilon_k}, \quad \epsilon_k = \sqrt{1+g^2-2g\cos k}.
\end{equation}
The diagonalized Hamiltonian is
\begin{equation}
    \hat{H}=\sum_k \epsilon_k\left(2 \hat{\gamma}_k^{\dagger} \hat{\gamma}_k-1\right), 
\end{equation}
and its ground state is given by
\begin{equation}
\left|\psi(g)\right\rangle = \prod_{0 < k <\pi }\left(\cos{\left(\frac{\theta_k}{2}\right)}-\sin {\left(\frac{\theta_k}{2}\right)} \hat{c}_k^{\dagger} \hat{c}_{-k}^{\dagger}\right)|0\rangle,
\end{equation}
where $\ket{0} = |\rightarrow\rightarrow\ldots\rightarrow \rangle$ denotes the fermion vacuum and the product is over all $0<k<\pi$ in Eq. \eqref{eq:wavenos}.

\subsection{Conventions for RTFIM diagonalization}
We now recall the Nambu formalism for constructing the ground state of the random transverse-field Ising model\cite{mbeng2020quantum}, Eq. \eqref{eq:rfim}. For simplicity, we will restrict our attention to even $N$ and even parity $\hat{P}$. The random transverse-field Ising Hamiltonian can be written in terms of a ``Nambu spinor'' $\boldsymbol{\Psi}$ as 
\begin{equation} \label{Nambu1}
\boldsymbol{\Psi}=\begin{pmatrix}
\hat{c}_1 \\
\vdots \\
\hat{c}_N \\
\hat{c}_1^{\dagger} \\
\vdots \\
\hat{c}_N^{\dagger}
\end{pmatrix}=\begin{pmatrix}
\hat{\mathbf{c}} \\
\hat{\mathbf{c}}^{\dagger}
\end{pmatrix}, \quad \hat{H} =\boldsymbol{\Psi}^{\dagger} \mathbf{H} \boldsymbol{\Psi}=\begin{pmatrix}\hat{\mathbf{c}}^{\dagger} & \hat{\mathbf{c}}\end{pmatrix}\begin{pmatrix}
\mathbf{A} & \mathbf{B} \\
-\mathbf{B} & -\mathbf{A}
\end{pmatrix}\begin{pmatrix}
\hat{\mathbf{c}} \\
\hat{\mathbf{c}}^{\dagger}
\end{pmatrix}
\end{equation}
where $\mathbf{H}$ is a $c$-number matrix to be diagonalized. By Hermiticity, $\mathbf{A}$ is symmetric and $\mathbf{B}$ is anti-symmetric, and the choice
\begin{equation}
\left\{\begin{array} { l }
{ \mathbf { A } _ { j , j } = g_ { j } } \\
{ \mathbf { A } _ { j , j + 1 } = \mathbf { A } _ { j + 1 , j } = - \frac {1} { 2 }}
\end{array}\right. , \quad
\left\{\begin{array}{l}
\mathbf{B}_{j, j}=0 \\
\mathbf{B}_{j, j+1}=-\mathbf{B}_{j+1, j}=-\frac{1}{2}
\end{array}\right.
\end{equation}
recovers Eq. \eqref{eq:rfim}. 
By even parity  $\mathbf{A}_{N,1} = \mathbf{A}_{1,N} = 1/2$ and $\mathbf{B}_{N,1} = -\mathbf{B}_{1,N} = 1/2$. In the following we try to find a basis transformation of these fermionic operators to diagonalise the Hamiltonian, while preserving the fermionic anti-commutation relations. Here $\mathbf{H}$ is a real anti-symmetric matrix and can be diagonalized by a unitary matrix. Let us write the spectral problem for $\mathbf{H}$ as
\begin{equation}
\mathbf{H}\left(\begin{array}{c}
\mathbf{u}_\mu \\
\mathbf{v}_\mu
\end{array}\right)=\left(\begin{array}{cc}
\mathbf{A} & \mathbf{B} \\
-\mathbf{B}^* & -\mathbf{A}^*
\end{array}\right)\left(\begin{array}{l}
\mathbf{u}_\mu \\
\mathbf{v}_\mu
\end{array}\right)=\epsilon_\mu\left(\begin{array}{l}
\mathbf{u}_\mu \\
\mathbf{v}_\mu
\end{array}\right)
\end{equation}
Note that if $\begin{pmatrix}\mathbf{u}_\mu & \mathbf{v}_\mu\end{pmatrix}^{\mathrm{T}}$ is eigenvector with eigenvalue $\epsilon_\mu$, then $\begin{pmatrix} \mathbf{v}_\mu^* & \mathbf{u}_\mu^*\end{pmatrix}^{\mathrm{T}}$ is an eigenvector with eigenvalue $-\epsilon_\mu$. Taking these eigenvectors to be pairwise orthonormal, a unitary change of basis is given by
\begin{equation}
\mathbf{Q}=\left(\begin{array}{ccc|ccc}
\mathbf{u}_1 & \cdots & \mathbf{u}_N & \mathbf{v}_1^* & \cdots & \mathbf{v}_N^* \\
\hline \mathbf{v}_1 & \cdots & \mathbf{v}_N & \mathbf{u}_1^* & \cdots & \mathbf{u}_N^*
\end{array}\right)=\left(\begin{array}{cc}
\mathbf{U} & \mathbf{V}^* \\
\mathbf{V} & \mathbf{U}^*
\end{array}\right),
\end{equation}
and brings $\mathbf{H}$ to the form $\mathbf{E} = \mathbf{Q^\dagger HQ} = \operatorname{diag}(\epsilon_1,\epsilon_2,\ldots,\epsilon_N, - \epsilon_1,-\epsilon_2,\ldots,-\epsilon_N)$. We now define a new set of fermions
\begin{equation}
\label{trans}
  \mathbf{\Phi}=\begin{pmatrix}
\hat{\vec{\gamma}} \\
\hat{\vec{\gamma}}^{\dagger}
\end{pmatrix}=\mathbf{Q}^{\dagger} \mathbf{\Psi}=\begin{pmatrix}
\mathbf{U}^{\dagger} & \mathbf{V}^{\dagger} \\
\mathbf{V}^{\mathrm{T}} & \mathbf{U}^{\mathrm{T}}
\end{pmatrix}\begin{pmatrix}
\hat{\mathbf{c}} \\
\hat{\mathbf{c}}^{\dagger}
\end{pmatrix}
\end{equation}
so that
\begin{equation}
  \hat{H} = \boldsymbol{\Psi}^{\dagger} \mathbf{H} \boldsymbol{\Psi} = \boldsymbol{\Phi}^{\dagger} \mathbf{Q^\dagger H Q} \boldsymbol{\Phi} = \boldsymbol{\Phi}^{\dagger} \mathbf{E} \boldsymbol{\Phi}.
\end{equation}
In this way, we have diagonalized $\hat{H}$ and the Hamiltonian can be written explicitly as
\begin{equation} 
\hat{H}=\sum_{\mu=1}^N\epsilon_\mu \left(2\hat{\gamma}_\mu^{\dagger} \hat{\gamma}_\mu-1\right).
\end{equation}

We now solve for ground states of this Hamiltonian of the form\cite{mbeng2020quantum}
\begin{equation} \label{ansatz}
  \ket{\psi(\mathbf{g})} = \mathcal{N} \exp \left(\frac{1}{2} \sum_{j_1 j_2} \mathbf{Z}_{j_1 j_2} \hat{c}_{j_1}^{\dagger} \hat{c}_{j_2}^{\dagger}\right)|0\rangle
  := \mathcal{N} e^{\hat{\zeta}} \ket{0}
\end{equation}
The ground state must satisfy the constraint $\hat{\gamma}_\mu \ket{\psi(\mathbf{g})} = 0 \ \forall \mu$, and we therefore require that
\begin{equation}
\sum_{j=1}^N\left(\mathbf{U}_{j \mu}^* \hat{c}_j+\mathbf{V}_{j \mu}^* \hat{c}_j^{\dagger}\right) e^{\hat{\zeta}} \ket{0}= 0.
\end{equation}
These conditions are solved by taking $\mathbf{Z} = -(\mathbf{U}^\dagger)^{-1}\mathbf{V^\dagger}$, and there further exists a unitary matrix $\mathbf{D}$ such that $\mathbf{Z = D\Lambda D}^\mathrm{T}$ and\cite{mbeng2020quantum}
\begin{equation}
  \boldsymbol{\Lambda}=\left(\begin{array}{cc|cc|c}
0 & \lambda_1 & 0 & 0 & \cdots \\
-\lambda_1 & 0 & 0 & 0 & \cdots \\
\hline 0 & 0 & 0 & \lambda_2 & \cdots \\
0 & 0 & -\lambda_2 & 0 & \cdots \\
\hline \vdots & \vdots & \vdots & \vdots & \vdots
\end{array}\right).
\end{equation}
Define new fermions $\mathbf{d^\dagger = D^\mathrm{T} \hat{c}^\dagger}$. Labelling consecutive columns of $\mathbf{D}$ as $1, \bar{1}; 2, \bar{2}, $etc, we can then write
\begin{equation} \label{result}
\ket{\psi(\mathbf{g})}=\mathcal{N} \exp \left(\sum_{p=1}^{N / 2} \lambda_p \hat{d}_p^{\dagger} \hat{d}_{\bar{p}}^{\dagger}\right)|0\rangle=\mathcal{N} \prod_{p=1}^{N / 2}\left(1+\lambda_p \hat{d}_p^{\dagger} \hat{d}_{\bar{p}}^{\dagger}\right)|0\rangle
\end{equation}
A suitable normalisation constant can be calculated using
\begin{equation}
  1 = \braket{\psi(\mathbf{g})|\psi(\mathbf{g})} = \mathcal{N}^2 \prod_{p=1}^{N/2} (1+|\lambda_p|^2) \longrightarrow \mathcal{N} = \sqrt{|\operatorname{det}(\mathbf{U})|}.
\end{equation}
In terms of the eigenvalues $\lambda_p$,
\begin{equation}
  \ket{\psi(\mathbf{g})}=\prod_{p=1}^{N / 2} \frac{1}{\sqrt{1+\left|\lambda_p\right|^2}}\left(1+\lambda_p \hat{d}_p^{\dagger} \hat{d}_{\bar{p}}^{\dagger}\right)|0\rangle=\prod_{p=1}^{N / 2}\left(u_p+v_p \hat{d}_p^{\dagger} \hat{d}_{\bar{p}}^{\dagger}\right)|0\rangle
\end{equation}
where we have defined $u_p=1 / \sqrt{1+\left|\lambda_p\right|^2}$ and $v_p=\lambda_p / \sqrt{1+\left|\lambda_p\right|^2}$,

From here we can immediately calculate the overlap between the ground state and the fermionic vacuum $\ket{0}$,
\begin{equation} \label{overlap}
  \braket{0|\psi(\mathbf{g})} = \bra{0} \mathcal{N} \prod_{p=1}^{N / 2}\left(1+\lambda_p \hat{d}_p^{\dagger} \hat{d}_{\bar{p}}^{\dagger}\right)\ket{0} = \mathcal{N} = \sqrt{|\operatorname{det}(\mathbf{U})|}.
\end{equation}

To use the random transverse-field Ising ground state in Eq. \eqref{eq:theorem1}, it remains to compute the overlap between this state and the GHZ state.

To this end, let us solve for the random transverse-field Ising ground state as an excited state of a transverse-field Ising ground state $|\psi(g)\rangle$, with no relation assumed for now between $\mathbf{g}$ and $g$. Specifically, just as in \eqref{ansatz} we solved for the ground state in terms of the fermionic vacuum $\ket{0}$ and corresponding fermion creation operators $\{c^\dagger_j\}$, we can equivalently solve for the ground state in terms of the Bogoliubov vacuum $|\psi(g)\rangle$ and the associated Bogoliubov operators $\{\gamma^\dagger_\mu\}$.

Thus we seek $\hat{Y}$ such that
\begin{equation}
  \ket{\psi(\mathbf{g})} = \mathcal{N} \exp \left(\frac{1}{2} \sum_{\mu_1 \mu_2} \mathbf{Y}_{\mu_1 \mu_2} \hat{\gamma}_{\mu_1}^{\dagger} \hat{\gamma}_{\mu_2}^{\dagger}\right)\ket{\psi(g)} := \mathcal{N} e^{\hat{Y}} \ket{\psi(g)}
\end{equation}

Denoting the quasiparticle operators for
$\ket{\psi(\mathbf{g})}$ by $\boldsymbol{\eta}^\dagger$, the two sets of fermionic quasiparticle creation operators $\boldsymbol{\eta}^\dagger$ and $\boldsymbol{\gamma}^\dagger$ are each related to the original Jordan-Wigner fermions $\boldsymbol{c}^\dagger$ via unitary transformations of the form
\begin{equation}
  \left(\begin{array}{l}
  \boldsymbol{\gamma} \\
  \boldsymbol{\gamma}^{\dagger}
  \end{array}\right)=
  \mathbf{Q_\gamma^\dagger}\left(\begin{array}{l}
  \hat{\mathbf{c}} \\
  \hat{\mathbf{c}}^{\dagger}
  \end{array}\right)=
  \left(\begin{array}{ll}
  \mathbf{U_\gamma}^{\dagger} & \mathbf{V_\gamma}^{\dagger} \\
  \mathbf{V_\gamma}^{\mathrm{T}} & \mathbf{U_\gamma}^{\mathrm{T}}
  \end{array}\right)\left(\begin{array}{l}
  \boldsymbol{c} \\
  \boldsymbol{c}^{\dagger}
  \end{array}\right)
\end{equation}

\begin{equation}
  \left(\begin{array}{l}
  \boldsymbol{\eta} \\
  \boldsymbol{\eta}^{\dagger}
  \end{array}\right)=
  \mathbf{Q_\eta^\dagger}\left(\begin{array}{l}
  \hat{\mathbf{c}} \\
  \hat{\mathbf{c}}^{\dagger}
  \end{array}\right)=
  \left(\begin{array}{ll}
  \mathbf{U_\eta}^{\dagger} & \mathbf{V_\eta}^{\dagger} \\
  \mathbf{V_\eta}^{\mathrm{T}} & \mathbf{U_\eta}^{\mathrm{T}}
  \end{array}\right)\left(\begin{array}{l}
  \boldsymbol{c}  \\
  \boldsymbol{c}^{\dagger}
  \end{array}\right)
\end{equation}
This immediately yields a unitary transformation relating $\boldsymbol{\eta}^\dagger$ and $\boldsymbol{\gamma}^\dagger$:
\begin{equation}
  \left(\begin{array}{l}
  \boldsymbol{\eta} \\
  \boldsymbol{\eta}^{\dagger}
  \end{array}\right)=
  \mathbf{Q_\eta^\dagger}\mathbf{Q_\gamma}\left(\begin{array}{l}
  \hat{\mathbf{\gamma}} \\
  \hat{\mathbf{\gamma}}^{\dagger}
  \end{array}\right)
\end{equation}
By analogy with Eq. \eqref{trans}, we define $\mathbb{Q}^\dagger := \mathbf{Q_\eta^\dagger}\mathbf{Q_\gamma}$, such that
\begin{equation}
\mathbb{Q} \equiv\left(\begin{array}{ll}
\mathbb{U} & \mathbb{V}^* \\
\mathbb{V} & \mathbb{U}^*
\end{array}\right),
\end{equation}
where
\begin{equation}
\mathbb{U}=\mathbf{U}_\gamma^{\dagger} \mathbf{U}_\eta+\mathbf{V}_\gamma^{\dagger} \mathbf{V}_\eta,
\quad
\mathbb{V}=\mathbf{V}_\gamma^{\mathrm{T}} \mathbf{U}_\eta+\mathbf{U}_\gamma^{\mathrm{T}} \mathbf{V}_\eta.
\end{equation}
Thus, by analogy with Eq. \eqref{overlap}, it follows that
\begin{equation}
\label{eq:RTFIMOverlap}
  |\braket{\psi(g)|\psi(\mathbf{g})}|^2 = |\operatorname{det}(\mathbb{U})|.
\end{equation}
By taking $g \to 0^+$, this result allows for direct computation of the quantum winning probability Eq. \eqref{eq:theorem1} for random transverse-field Ising ground states $|\psi(\mathbf{g})\rangle$ through manipulating $2N$-by-$2N$ matrices (rather than diagonalizing $2^N$-by-$2^N$ matrices as would be needed for a brute-force computation).
\section{Perturbative calculations} \label{ap:2ndpt}
We now consider the random transverse-field Ising Hamiltonian as a weak perturbation of the transverse-field Ising model, to wit
\begin{equation}
\hat{H} = \hat{H}_{\mathrm{TFIM}} + \delta \hat{H},
\end{equation}
with $\hat{H}_{\mathrm{TFIM}}$ given by Eq. \eqref{eq:TFIM} and
\begin{equation} \label{eq:pert}
    \delta H = -\sum_{j=1}^N \delta g_j \hat{X}_j = \sum_{j=1}^N \delta g_j\left( 2 \hat{c}_j^\dagger \hat{c}_j - 1 \right) 
\end{equation}
in terms of Jordan-Wigner fermions. Our goal is to calculate the change in the quantum winning probability Eq. \eqref{eq:explicit_pqu}, i.e. in the wavefunction overlap $|\braket{\mathrm{GHZ}^+|\psi(\mathbf{g})}|^2$, given that $|\psi(\mathbf{g})\rangle$ is a ground state of the random transverse-field Ising model with weak randomness $g_j = g + \delta g_j$ and $|\delta g_j|\ll g$. We will specifically be concerned with terms up to second order in the perturbation strength (cf. the discussion in Sec. \ref{sec:gentheor}). To this order, the even-parity case of Eq. \eqref{eq:theorem1} can be expanded perturbatively in the state $|\psi\rangle$ to yield
\begin{equation}
    \begin{aligned}
        p_\mathrm{qu}(\ket{\psi+\delta \psi}) &= \frac{1}{2} + \frac{1}{2}\left|\braket{\mathrm{GHZ}^+|\psi}+\braket{\mathrm{GHZ}^+|\delta\psi^{(1)}}+\braket{\mathrm{GHZ}^+|\delta\psi^{(2)}}\right|^2 + \mathcal{O}((\delta g)^3) \\
        & = \underbrace{ \frac{1}{2} + \frac{1}{2}|\braket{\mathrm{GHZ}^+|\psi}|^2}_{p_\mathrm{qu}(\ket{\psi})} + \underbrace{\braket{\mathrm{GHZ}^+|\psi}\braket{\mathrm{GHZ}^+|\delta\psi^{(1)}}}_{\delta p_\mathrm{qu}^{(1)}\text{, first variation}}+\underbrace{\frac{1}{2}\braket{\mathrm{GHZ}^+|\delta\psi^{(1)}}^2+\braket{\mathrm{GHZ}^+|\psi}\braket{\mathrm{GHZ}^+|\delta\psi^{(2)}}}_{\delta p_\mathrm{qu}^{(2)}\text{, second variation}}
    \end{aligned}.
\end{equation}
In the remainder of this Appendix, we present exact expressions for the first and second variation of $p_{\mathrm{qu}}(|\psi\rangle)$, given that $|\psi\rangle$ is perturbed by the fields $\delta g_j$ according to Eq. \eqref{eq:pert}.

\subsection{The first variation of \(p_{\mathrm{qu}}\)}
To calculate the first variation of $p_{\mathrm{qu}}$, we must first obtain the first-order perturbed state $\ket{\delta \psi}$. By first-order perturbation theory,
\begin{equation} \label{eq:1stOrderPT}
    \braket{\text{GHZ}^+ |\delta \psi} = \sum_{\vec{n} \in \{0,1\}^N} \frac{\bra{\psi_{\vec{n}}(g)}\delta \hat{H} \ket{\psi(g)}}{E_{\vec{0}}-E_{\vec{n}}} \braket{\text{GHZ}^+ |\psi_{\vec{n}}(g)}.
\end{equation}
where the bit string $\vec{n} = (n_1,n_2,\ldots,n_N)$ labels the excited states of the TFIM, $\ket{\psi_{\vec{n}}(g)} = \prod_{k} \left(\hat{\gamma}_k^\dagger \right)^{n_k} |\psi(g)\rangle$. 

We observe that $\delta H$ does not change the fermionic parity of the state, so only terms with even $\sum_k n_k$ contribute to this sum. First consider the numerator in Eq. \eqref{eq:1stOrderPT},
\begin{equation} \label{eq:Overlap}
    \bra{\psi_{\vec{n}}} \delta H \ket{\psi(g)} = \frac{2}{N} \sum_j \delta g_j \sum_{k, k' } e^{-i(k-k')j}\bra{\psi_{\vec{n}}(g)} \hat{c}_k^\dagger \hat{c}_{k'}  \ket{\psi(g)}.
\end{equation}
In terms of Bogoliubov quasiparticle operators, this is given by
\begin{equation} \label{eq:factor1}
  \begin{aligned}
      \bra{\psi_{\vec{n}}(g)} \hat{c}_k^\dagger \hat{c}_{k'}  \ket{\psi(g)} &= \bra{\psi(g)} \prod_{q} (\hat{\gamma}_q)^{n_q} \left(\cos \frac{\theta_k}{2} \hat{\gamma}_k^{\dagger}-\sin \frac{\theta_k}{2} \hat{\gamma}_{-k}\right)\left(\cos \frac{\theta_{k^{\prime}}}{2} \hat{\gamma}_{k^{\prime}}-\sin \frac{\theta_{k^{\prime}}}{2} \hat{\gamma}_{-k^{\prime}}^{\dagger}\right) \ket{\psi(g)}. \\
      &= -\cos \frac{\theta_k}{2} \sin \frac{\theta_{k'}}{2} \bra{\psi(g)}\hat{\gamma}_q\hat{\gamma}_{q'} \hat{\gamma}_k^\dagger\hat{\gamma}_{-k'}^\dagger \ket{\psi(g)} \\
    &=  -\cos \frac{\theta_k}{2} \sin \frac{\theta_{k'}}{2} \left( \delta_{q'k}\delta_{q,-k'} - \delta_{q',-k'}\delta_{qk} \right)
  \end{aligned}
\end{equation}
In the second equality above, we restricted our attention to $\ket{\psi_{\vec{n}}}=\ket{\psi_{qq'}} = \hat{\gamma}_q^\dagger\hat{\gamma}_{q'}^{\dagger} \ket{\psi(g)}$ i.e. virtual states with exactly two excitations, fixing the convention that $q<q'$. It remains to compute $\braket{\mathrm{GHZ}^+|\psi_{qq'}}$. For a non-zero overlap, $\ket{\psi_{qq'}}$ must respect the particle-hole symmetry of the GHZ state, and we may choose $q=-q' < 0$. We can express the excited states as\cite{mbeng2020quantum}
\begin{equation}
    \ket{\psi_{q'}}= -\left(\sin \frac{\theta_{q'}}{2}+\cos \frac{\theta_{q^{\prime}}}{2} \hat{c}_{q^{\prime}}^{\dagger} \hat{c}_{-q^{\prime}}^{\dagger}\right) \ket{\psi_0}.
\end{equation}
Thus
\begin{equation} \label{eq:factor2}
  \braket{\mathrm{GHZ}^+ | \psi_{q q^{\prime}}}=\braket{\psi(0^{+}) | \psi_{q q^{\prime}}} = -\left[\prod_{0<k<\pi} \cos \left(\frac{\theta_k-\theta_{k}^0}{2}\right)\right] \cdot \tan \left(\frac{\theta_{q^{\prime}}-\theta_{q^{\prime}}^0}{2}\right) \delta_{q,-q^{\prime}}
\end{equation}
where $\theta_k^0 := \theta_k(0^+)$. In Eq. \eqref{eq:1stOrderPT}, this yields
\begin{equation} \label{eq:factor2contd}
    \braket{\mathrm{GHZ}^+ |\delta\psi} = \frac{1}{2} \left(\frac{1}{N} \sum_{j=1}^N \delta g_j\right) \left(\sum_{0<k<\pi}\frac{1}{\epsilon_k}\tan \left(\frac{\theta_{k}^0-\theta_{k}}{2}\right) \sin \theta_k \right) \braket{\text{GHZ}^+|\psi(g)},
\end{equation}
and to first order in $\lambda$,
\begin{equation} \label{eq:dp1calc}
    \delta p_\mathrm{qu}^{(1)} = p_\mathrm{qu}(\ket{\psi + \delta \psi}) - p_\mathrm{qu}(\ket{\psi}) = \frac{|\braket{\text{GHZ}^+|\psi(g)}|^2}{2N}\sum_{0<k<\pi } -f_k(g) \sum_{j=1}^N \delta g_j,
\end{equation}
where $f_k$ is as defined in Eqn. \eqref{eq:fk}. To verify this result, note by the chain rule that
\begin{equation} \label{eq:dp1}
\delta p_\mathrm{qu}^{(1)} = \frac{(p_{\mathrm{qu}}-1/2)}{N} \left(\sum_{j=1}^N \delta g_j\right) \chi'(g),
\end{equation}
which implies that
\begin{equation}
\chi'(g) = \sum_{0<k<\pi } f_k(g), 
\end{equation}
which is consistent with Eq. \eqref{eq:d1chi}.

\subsection{The second variation of $p_{\mathrm{qu}}$}
The second variation of $p_{\mathrm{qu}}$ is rather more complicated to analyze. However, the method is analogous to the first-order calculation above, and we merely quote the final result:
\begin{equation} \label{eq:dp2}
\begin{aligned}
    & \delta p_\mathrm{qu}^{(2)} =\frac{p_\mathrm{qu}-1/2}{N^2} \sum_{p_1>0, p_2>0} f_{p_1}f_{p_2} \sum_{j,l=1}^N \delta g_j \delta g_l+\left\{\frac{p_\mathrm{qu}-1/2}{N^2}\sum_{\substack{p_2>0,p_1>0}} \frac{1}{(\epsilon_{p_1}+\epsilon_{p_2})^2} \tan \left(\frac{\theta_{p_1}^0 - \theta_{p_1}}{2}\right) \tan \left(\frac{\theta_{p_2}^0 - \theta_{p_2}}{2}\right) \right. \\
  &\times \left.   \sum_{j,l = 1}^N \delta g_j\delta g_l \left[\sin^2 \left(\frac{\theta_{p_2}-\theta_{p_1}}{2}\right) \cos \left((p_1+p_2)(j-l)\right) - \sin^2 \left(\frac{\theta_{p_2}+\theta_{p_1}}{2}\right) \cos \left((p_1-p_2)(j-l)\right)\right] \right\} \\
  & + \left\{\frac{p_\mathrm{qu}-1/2}{N^2} \sum_{j,l=1}^N \delta g_j \delta g_l \sum_{p_1>0, p_2>0} \frac{1}{2(\epsilon_{p_1}+\epsilon_{p_2})} \left( \sin (\theta_{p_2}-\theta_{p_1}) \left[\frac{f_{p_2}}{\sin \theta_{p_2}} - \frac{f_{p_1}}{\sin \theta_{p_1}}\right] \times \cos (p_1+p_2)(j-l)  \right. \right. \\
  & + \left.\left. \sin (\theta_{p_2}+\theta_{p_1}) \left[\frac{f_{p_2}}{\sin \theta_{p_2}} + \frac{f_{p_1}}{\sin \theta_{p_1}}\right] \times \cos (p_1-p_2)(j-l) \right)\right\} \\
  &  - \left\{ \frac{p_\mathrm{qu}-1/2}{N^2}  \sum_{j,l=1}^N \delta g_j \delta g_l \sum_{p_1>0, p_2>0} \frac{1}{(\epsilon_{p_1}+\epsilon_{p_2})^2} \times   \right. \\
  &  \times \left. \left( \sin^2 \left(\frac{\theta_{p_2}-\theta_{p_1}}{2}\right) \cos (p_1+p_2)(j-l) + \sin^2 \left(\frac{\theta_{p_2}+\theta_{p_1}}{2}\right) \cos (p_1-p_2)(j-l) \right) \right\}.
\end{aligned}
\end{equation}
To relate this to the utility function $u(\mathbf{g})$ of interest in the main text, note that by Eq. \eqref{eq:utilityfn},
\begin{equation}
    \delta^{(1)}u(\mathbf{g}) = \frac{\delta^{(1)} p_\mathrm{qu}}{p_\mathrm{qu}-p_r},
\end{equation}
from which it follows that
\begin{equation}\label{eq:relationdpdu}
    \delta^{(2)} p_\mathrm{qu} = \delta^{(1)}p_\mathrm{qu}\delta^{(1)} u(\mathbf{g}) + (p_\mathrm{qu}-p_r)\delta^{(2)} u(\mathbf{g}) = \frac{[\delta^{(1)}p_\mathrm{qu}]^2}{p_\mathrm{qu}-p_r} + (p_\mathrm{qu} - p_r) \delta^{(2)}u(\mathbf{g}).
\end{equation}
The first term in Eq. \eqref{eq:relationdpdu} matches the first term of \eqref{eq:dp2}, and $\delta u^{(2)}(g)$ can be found by summing over the rest of the terms in Eq. \eqref{eq:dp2} and dividing by the extra factor $p_\mathrm{qu}-p_r = \left| \braket{\mathrm{GHZ}^+|\psi(g=0^+)} \right|^2/2$.

We now show how this general formula yields simple expressions for $\nabla^2 u$ and $\chi''(g)$ in Eqs. \eqref{eq:d2chi} and \eqref{eq:nabla_u}. For perfectly correlated disorder $\langle \delta g_i \delta g_j\rangle = \delta g^2$, we have
\begin{equation}
    \sum_{j,l=1}^N \delta g_j \delta g_l \cos (p_1\pm p_2)(j-l) = \frac{1}{2}\sum_{j,l=1}^N e^{(p_1 \pm p_2)(j-l)} + \frac{1}{2}\sum_{j,l=1}^N e^{-(p_1 \pm p_2)(j-l)} = N^2 \delta g^2 \delta_{p_1,\mp p_2}.
\end{equation}
Since all the relevant momenta are positive, this eliminates terms with factors of $\cos(p_1+p_2)(j-l)$. The three terms in the braces can be simplified as
\begin{equation} \label{eq:ReproduceChi}
    \begin{aligned}
        \mathrm{Term \ 1} &= -\delta g^2 \sum_{p>0} \frac{1}{4 \epsilon_p^2} \tan^2 \left(\frac{\theta^0_{p}-\theta_{p}}{2}\right)\sin^2 \theta_p = -\sum_{p>0} \frac{1}{4} f_p^2, \\
        \mathrm{Term \ 2} &= \delta g^2\sum_{p>0} \frac{1}{4\epsilon_p} \sin 2\theta_p \times \frac{2f_p}{\sin \theta_p} = \sum_{p>0} \frac{f_p \sin \theta_p}{\epsilon_p}, \\
        \mathrm{Term \ 3} &= -\delta g^2\sum_{p>0} \frac{\sin^2 \theta_p}{4 \epsilon_p^2}.
    \end{aligned}
\end{equation}
Summing over all contributions yields 
\begin{equation}
    \delta u^{(2)} = \delta g^2 \left(\sum_{p>0} \frac{f_p \sin \theta_p}{\epsilon_p}- \frac{1}{4} f_p^2 - \frac{\sin^2 \theta_p}{4 \epsilon_p^2}\right) = \frac{1}{2} \chi''(g) \delta g^2,
\end{equation}
which recovers the expression in Eq. \eqref{eq:d2chi}. Next we turn to uncorrelated disorder where $\langle \delta g_i \delta g_j \rangle = \sigma^2 \delta_{ij}$. This gives
\begin{equation}
    \sum_{j,l=1}^N \sigma^2 \delta_{jl} \cos (p_1 \pm p_2)(j-l) = N\sigma^2
\end{equation}
\begin{equation}
    \begin{aligned}
        \mathrm{Term \ 1} &= \frac{\sigma^2}{N}\sum_{p_1,p_2>0} \frac{1}{(\epsilon_{p_1}+\epsilon_{p_2})^2} \tan \left(\frac{\theta_{p_1}^0 - \theta_{p_1}}{2}\right) \tan \left(\frac{\theta_{p_2}^0 - \theta_{p_2}}{2}\right) \sin \theta_{p_1} \sin \theta_{p_2} = -\frac{\sigma^2}{N}\sum_{p_1,p_2>0}  \frac{\epsilon_{p_1}\epsilon_{p_2}f_{p_1}f_{p_2}}{(\epsilon_{p_1}+\epsilon_{p_2})^2} \\
        \mathrm{Term \ 2} &= \frac{\sigma^2}{N} \sum_{p_1,p_2>0} \frac{1}{2(\epsilon_{p_1}+\epsilon_{p_2})} \left(\frac{f_{p_2}}{\sin \theta_{p_2}}\times 2\sin \theta_{p_2} \cos \theta_{p_1} + p_1 \leftrightarrow p_2 \right) = \frac{\sigma^2}{N}\sum_{p_1,p_2>0} \frac{f_{p_1}\cos \theta_{p_2}+f_{p_2}\cos \theta_{p_1}}{\epsilon_{p_1}+\epsilon_{p_2}} \\
        \mathrm{Term \ 3} &= -\frac{\sigma^2}{N}\sum_{p_1,p_2>0} \frac{1}{(\epsilon_{p_1}+\epsilon_{p_2})^2} \left(\sin^2 \left(\frac{\theta_{q_2}-\theta_{q_1}}{2}\right)+ \sin^2 \left(\frac{\theta_{q_2}+\theta_{q_1}}{2}\right) \right) = - \frac{\sigma^2}{N} \sum_{p_1,p_2>0} \frac{1+\cos \theta_{p_1}\cos\theta_{p_2}}{(\epsilon_{p_1}+\epsilon_{p_2})^2}
    \end{aligned}
\end{equation}
Summing these contributions yields
\begin{equation}
    \delta u^{(2)} =  \frac{\sigma^2}{N}\sum_{p_1,p_2>0}  -\frac{\epsilon_{p_1}\epsilon_{p_2}f_{p_1}f_{p_2}}{(\epsilon_{p_1}+\epsilon_{p_2})^2} + \frac{f_{p_1}\cos \theta_{p_2}+f_{p_2}\cos \theta_{p_1}}{\epsilon_{p_1}+\epsilon_{p_2}} - \frac{1+\cos \theta_{p_1}\cos\theta_{p_2}}{(\epsilon_{p_1}+\epsilon_{p_2})^2} \equiv \frac{1}{2} \delta g^2\nabla^2 u
\end{equation}

This leads to the expression for $\nabla^2 u$ in Eq. \eqref{eq:nabla_u}.
\section{Critical scaling of the rescaled second variation for perfectly correlated disorder} \label{ap:scaling}

In this Appendix, we study the singular behaviour of the rescaled second variation given by Eq. \eqref{eq:SecondVarMaxCorr}, in the approach to the Ising critical point $g=g_c=1$ as $N,\,g \to \infty$. In order to perform this analysis, it will be helpful to first study the asymptotic behaviour as $N\to\infty$ of the sums
\begin{equation}
    S_1 = \sum_k \frac{1}{\sin{k/2}},  \quad S_2 = \sum_k \frac{1}{\sin^2{k/2}}
\end{equation}
where $k = \pi/N, 3\pi/N, ..., (N-1)\pi/N$ as in Eq. \eqref{eq:wavenos}.

To evaluate $S_1$, it is helpful to introduce a cutoff $M$ such that $1 \ll M \ll N^{1/2}$ and is $M$ odd. Summing up to this scale yields 
\begin{equation} \label{eq:FiniteSizeSum}
    \sum_{k=\pi/N}^{M\pi/N}  \frac{1}{\sin{k/2}} =  \sum_{k=\pi/N}^{M\pi/N} \frac{2}{k} + \mathcal{O}(M^2/N) = \frac{2N}{\pi} \sum_{n=1}^{\frac{M+1}{2}} \frac{1}{2n-1} + \mathcal{O}(M^2/N).
\end{equation}
Recalling that the harmonic series has the asymptotic form
\begin{equation}
    H_K := \sum_{n=1}^K \frac{1}{n} = \log K +\gamma +\mathcal{O}(K^{-1})
\end{equation}
where $\gamma \approx 0.5772\dots$ denotes the Euler-Mascheroni constant, the sum in \eqref{eq:FiniteSizeSum} can be evaluated as
\begin{equation}
    \frac{N}{\pi} \left(H_{M+1}-\frac{1}{2}H_{\frac{M+1}{2}} \right) = \frac{N}{\pi}\log (M+1) + \frac{\gamma+\log{2}}{\pi}N + \mathcal{O}(M^{-1}).
\end{equation}
Next consider allowed values of $k\in [(M+1)\pi /N, \pi]$. The sum over this range of wavevectors is sufficiently far away from the singularity at $k=0$ that it can be safely be replaced by an integral, since 
\begin{equation} \label{eq:FiniteSizeInt}
    \begin{aligned}
        \sum_{k = \frac{M+1}{N}\pi}^{\frac{N-1}{N}\pi} \frac{1}{\sin \frac{k}{2}} &= \left(\frac{N}{2\pi}\right)\int_{\frac{M+1}{N}\pi}^\pi  \frac{dk}{\sin \frac{k}{2}} + \mathcal{O}(M^{-1}) \\
        & = \frac{N}{2\pi} \left(2\log \left(\tan \frac{k}{4}\right)\right)_{k=\frac{M+1}{N}\pi}^{k=\pi} + \mathcal{O}(M^{-1})\\
        & = \frac{1}{\pi} N \log N - \frac{N}{\pi}\log(M+1) - \frac{N}{\pi}\log \frac{\pi}{4}+ \mathcal{O}(M^{-1},M^2/N).
    \end{aligned}
\end{equation}
Combining Eqs. \eqref{eq:FiniteSizeSum} and \eqref{eq:FiniteSizeInt} and passing to the limit $M,\,N \to \infty$ with the cutoff $M/N^{1/2} \to 0$, we find that the leading asymptotic behaviour of $S_1$ is given by
\begin{equation}
\label{eq:s1}
    S_1 \sim \frac{1}{\pi} N \log N + \frac{N}{\pi}\left(\gamma + \log \frac{8}{\pi}\right).
\end{equation}
We now turn to $S_2$. It is again useful to introduce a cutoff $M$ that now satisfies $1 \ll M \ll N$. For allowed values of $k\in [\pi/N,M\pi/N]$, we have by similar reasoning to above,
\begin{equation}
    \sum_{k = \pi/N}^{M\pi/N} \frac{1}{\sin^2 \frac{k}{2}} = 4\sum_{k = \pi/N}^{M\pi/N} \frac{1}{k^2} + \mathcal{O}(M) 
    = \frac{4N^2}{\pi^2} \left(\frac{\pi^2}{8}-\frac{1}{2M}\right) + \mathcal{O}(M,N^2/M^2),
\end{equation}
where in the last step we used the Euler-Maclaurin formula for partial sums of $\zeta(2)$. Meanwhile for allowed values of $k\in [\frac{M+1}{N}\pi,\pi]$, we have
\begin{equation}
    \sum_{k=M\pi/N}^{\frac{N-1}{N}\pi} \frac{1}{\sin^2 \frac{k}{2}} 
    = \frac{N}{\pi} \cot \frac{M\pi}{2N} + \mathcal{O}(N/M^2) 
    = \frac{2N^2}{\pi^2 M} + \mathcal{O}(N/M^2,M).
\end{equation}
Combining these expressions, we deduce that the leading asymptotic behaviour of $S_2$ as $N \to \infty$ is given by
\begin{equation}
\label{eq:s2}
    S_2 = \sum_{k=\pi/N}^{\frac{N-1}{N}\pi} \frac{1}{\sin^2 \frac{k}{2}} \sim \frac{N^2}{2}.
\end{equation}
\subsection{Finite-size scaling at the critical point}

We are now in a position to derive the leading finite-size scaling of $\chi'(g_c)$ and $\chi''(g_c)$ in the large-system limit. We first note that
\begin{equation} \label{eq:fk1}
  f_k(g=g_c)= \frac{1}{2}\left(\frac{1}{\sin k/2}-1\right).
\end{equation}
Using Eq. \eqref{eq:s1},
\begin{equation}
    \chi'(g_c) = \sum_{k>0} -f_k(g=g_c) \sim -\frac{1}{2\pi}N\log N - \frac{N}{2\pi}\left(\gamma+ \log \frac{8}{\pi}\right) + \frac{N}{4}
\end{equation}
as $N \to \infty$. Similarly, $\chi''(g_c)$ can be evaluated at criticality as
\begin{equation}
    \chi''(g_c) =  \sum_{k=\pi/N}^{\frac{N-1}{N}\pi} -\frac{1}{4\sin^2 \frac{k}{2}}+\frac{3}{4\sin{\frac{k}{2}}}-\frac{1}{2} = -\frac{S_2}{4} + \frac{3S_1}{4} - \frac{N}{4}.
\end{equation}
It follows by Eqs. \eqref{eq:s1} and \eqref{eq:s2} that
\begin{equation}
    \chi''(g_c) \sim -\frac{N^2}{8}
\end{equation}
and thus the rescaled second variation in Eq. \eqref{eq:SecondVarMaxCorr} diverges linearly in $N$
\begin{equation}
    \frac{\delta u^{(2)}(g_c,g_c,\ldots,g_c)}{N\sigma^2} \sim - \frac{N}{16}
\end{equation}
at the critical point as the system size $N \to \infty$.

\subsection{Critical scaling in the large-system limit}
We next consider the critical scaling as a function of $g$ in the large system limit $N \gg 1$. It follows by Eqs. \eqref{eq:defb2} and \eqref{eq:defbIsing} that
\begin{equation} \label{eq:dchidg@therm}
  \begin{aligned}
      \frac{\mathrm{d}\chi}{\mathrm{d}g} &\sim -\frac{N}{2\pi}\int_{0}^{\pi} \frac{g\sin^2 k (1+g^2-2g\cos k)^{-1}}{\sqrt{1+g^2-2g\cos k}+1-g\cos k} \mathrm{d}k \sim \frac{N}{2\pi} \left\{ \frac{2}{g (g+1)} K\left(\frac{4 g}{(g+1)^2}\right) -\theta(1-g)\frac{\pi}{g} \right\}
  \end{aligned}
\end{equation}
as $N \to \infty$. Similarly for the second derivative,
\begin{equation} \label{eq:d2chi@therm}
    \frac{\mathrm{d}^2\chi}{\mathrm{d}g^2} \sim \frac{N}{2\pi} \left\{ \frac{1}{g^2(g-1)}E\left(\frac{4 g}{(g+1)^2}\right) + \frac{3}{g^2(g+1)}K\left(\frac{4 g}{(g+1)^2}\right) - \theta(1-g)\frac{\pi}{g^2}\right\}.
\end{equation}
where $\theta(x)$ denotes the Heaviside step function and $K$ and $E$ denote the elliptical integrals of the first and second kind respectively. Expanding Eq. \eqref{eq:d2chi@therm} near $g=g_c=1$ yields
\begin{equation}
    \frac{\delta u^{(2)}}{N\sigma^2} \sim \frac{1}{2(g-1)} - \frac{3}{4} \log |1-g|
\end{equation}
to leading order in $g-1$. The leading $(g-1)$-dependence of the rescaled second variation is the same on both sides of the critical point.
\end{document}